\documentclass[pdflatex,sn-chicago]{sn-jnl}

\usepackage{graphicx}%
\usepackage{multirow}%
\usepackage{amsmath,amssymb,amsfonts}%
\usepackage{amsthm}%
\usepackage{mathrsfs}%
\usepackage[title]{appendix}%
\usepackage{xcolor}%
\usepackage{textcomp}%
\usepackage{manyfoot}%
\usepackage{booktabs}%
\usepackage{algorithm}%
\usepackage{algorithmicx}%
\usepackage{algpseudocode}%
\usepackage{listings}%

\usepackage{graphicx}%
\usepackage{multirow}%
\usepackage{amsmath,amssymb,amsfonts}%
\usepackage{amsthm}%
\usepackage{mathrsfs}%
\usepackage[title]{appendix}%
\usepackage{xcolor}%
\usepackage{textcomp}%
\usepackage{manyfoot}%
\usepackage{booktabs}%
\usepackage{algorithm}%
\usepackage{algorithmicx}%
\usepackage{algpseudocode}%
\usepackage{listings}%

\usepackage{multirow} 
\usepackage{booktabs} 
\usepackage{xcolor} 
\usepackage{colortbl} 
\usepackage{graphicx}
\usepackage{subcaption}
\usepackage{enumitem}
\usepackage{dialogue}
\usepackage{verbatim}

\usepackage{xcolor}
\usepackage{colortbl}
\definecolor{gray}{gray}{0.9}
\definecolor{orange}{rgb}{1,0.5,0}
\definecolor{mdred}{rgb}{0.7,0,0}
\definecolor{mdgreen}{rgb}{0.05,0.6,0.05}
\definecolor{mdblue}{rgb}{0,0,0.7}
\definecolor{dkblue}{rgb}{0,0,0.5}
\definecolor{dkgray}{rgb}{0.3,0.3,0.3}
\definecolor{slate}{rgb}{0.25,0.25,0.4}
\definecolor{ltgray}{rgb}{0.7,0.7,0.7}
\definecolor{purple}{rgb}{0.7,0,1.0}
\definecolor{lavender}{rgb}{0.65,0.55,1.0}

\newcommand{\amr}[1]{\texttt{#1}}

\usepackage{adjustbox,booktabs,tabularx,ragged2e}

\newcommand{\Time}[1]{{#1}}
\newcommand{\Aspect}[1]{{#1}}

\definecolor{red}{rgb}{0,0,0}
\newcommand{\new}[1]{\textcolor{red}{#1}}
\usepackage{soul}

\theoremstyle{thmstyleone}%
\theoremstyle{thmstyletwo}%

\theoremstyle{thmstylethree}%

\begin{document}

\title[Human-Robot Dialogue Annotation for Multi-Modal Common Ground]{Human-Robot Dialogue Annotation for Multi-Modal Common Ground}


\author*[1]{\fnm{Claire} \sur{Bonial}}\email{claire.n.bonial.civ@army.mil}
\equalcont{These authors contributed equally to this work.}

\author*[1]{\fnm{Stephanie M.} \sur{Lukin}} 
\email{stephanie.m.lukin.civ@army.mil}
\equalcont{These authors contributed equally to this work.
}

\author[1]{\fnm{Mitchell} \sur{Abrams}}
\author[1]{\fnm{Anthony} \sur{Baker}}
\author[2]{\fnm{Lucia} \sur{Donatelli}}
\author[1]{\fnm{Ashley} \sur{Foots}}
\author[1]{\fnm{Cory J.} \sur{Hayes}}
\author[1]{\fnm{Cassidy} \sur{Henry}}
\author[3]{\fnm{Taylor} \sur{Hudson}}
\author[4]{\fnm{Matthew} \sur{Marge}}
\author[1]{\fnm{Kimberly A.} \sur{Pollard}} 
\author[5]{\fnm{Ron} \sur{Artstein}}
\author[5]{\fnm{David} \sur{Traum}}
\author[1]{\fnm{Clare R.} \sur{Voss}}

\affil[1]{DEVCOM Army Research Laboratory, Adelphi MD, USA}


\affil[2]{Vrije Universiteit, Amsterdam, The Netherlands}


\affil[3]{Oak Ridge Associated Universities, Oak Ridge, TN, USA}

\affil[4]{DARPA, USA}

\affil[5]{USC Institute for Creative Technologies, Playa Vista, CA, USA}


\abstract{In this paper, we describe the development of symbolic representations annotated on human-robot dialogue data to make dimensions of meaning accessible to autonomous systems participating in collaborative, natural language dialogue, and to enable {\it common ground} with human partners. 
\new{A particular challenge for establishing common ground arises in remote dialogue (occurring in disaster relief or search-and-rescue tasks), where a human and robot are engaged in a joint navigation and exploration task of an unfamiliar environment, but where the robot cannot immediately share high quality visual information due to limited communication constraints. Engaging in a dialogue provides an effective way to communicate, while on-demand or lower-quality visual information can be supplemented for establishing common ground.
Within this paradigm, we capture propositional semantics and the illocutionary force of a single utterance within the dialogue through our Dialogue-AMR annotation, an augmentation of Abstract Meaning Representation.
We then capture patterns in how different utterances within and across speaker floors relate to one another in our development of a multi-floor Dialogue Structure annotation schema.
Finally, we begin to annotate and analyze the ways in which the visual modalities provide contextual information to the dialogue for overcoming disparities in the collaborators' understanding of the environment.}
We conclude by discussing the use-cases, architectures, and systems we have implemented from our annotations that enable physical robots to autonomously engage with humans in bi-directional dialogue and navigation.}


\maketitle

\section{Introduction}
\label{intro}

What computational language resources are needed for robots to act as partners to humans 
in disaster relief and search-and-rescue tasks? There exist multiple scenarios where they can act as scouts exploring an area while their human partners gain an understanding of the situation 
on the ground and also remain safely at a remote location. 
Far from being the stuff of science fiction, robots are already being leveraged 
effectively in disaster response \new{and robot rescue  \citep{nagatani2013emergency,murphy2014disaster,habibian2021design,edlinger2022intuitive,kanazawa12023considerations}.}
However, the current state of the art largely relies upon human tele-operation of the robot using a handheld controller (i.e. no autonomy) \new{(e.g., \cite{kang2003robhaz,ryu2004multi,yamauchi2004packbot,chiou2022robot})}.\footnote{\new{A meta-analysis of multi-agent systems for search and rescue applications points out that of 40 documented cases of robot-assisted disaster response, only two involved a degree of autonomous navigation in marine robots \citep{drew2021multi}.}}
Less frequently, some approaches have leveraged an initial static tasking of the robot (e.g., providing the robot with a specific point in a pre-mapped area to which it should navigate) followed by autonomous path planning and navigation to move to that point (e.g., \cite{williams2012monitoring,arvidson2010spirit,camilli2010tracking}). 
These current approaches place the training and cognitive burden of robot operation on the human partner 
and do not facilitate \new{the flexible, real-time re-tasking needed in the dangerous and dynamic environments of disaster relief \citep{app13031800}.}

Large language models (LLMs) have increasingly been leveraged for natural language interaction with robots. Current approaches tend to involve the use of an LLM to effectively ``translate'' from natural language into an executable action of the robot (e.g., \citet{brohan2023can}). The major weaknesses of such approaches are first, that LLMs do not provide optimal or even feasible plans,\footnote{Recent extensive evaluations of LLMs in generating executable plans (\textit{sans robots}) found only an average success rate of 12 percent across domains by even the best model (GPT-4) \citep{valmeekam2023}.  However, other research has shown more promise when leveraging re-prompting for plan correction \citep{silver2024generalized,chen2024autotamp}.} which traditional AI planners are able to do, and second, LLMs are prone to factual inaccuracies and hallucinations.  Although very recent research has sought to overcome these weaknesses (described in related work), LLM architectures have not consistently demonstrated the ability to surmount these issues at the scale and to the degree required for disaster relief.

Furthermore, these approaches focus on one human tasking one robot with well-defined capabilities. There is a relatively unexplored opportunity for heterogeneous human-robot teams that would better address the needs of the disaster
relief domain, which requires team members to be physically distributed, often under dangerous conditions \citep{drew2021multi,queralta2020collaborative}. 
Under the current state of the art, the technical challenge of instructing heterogeneous teams of robots requires an additional human operator to tele-operate or provide an individualized task plan for each robot, relying on the human understanding of its capabilities measured against the risks and hazards of the environment. Additionally, such approaches preclude flexible combinations of capabilities of different robots that may not be obvious to operators or available in individual tasking. 

To enable flexible tasking of one or more robots with a low cognitive burden on operators,\footnote{The objective is to limit the number and mix of robots and tasks that the operators in an unknown environment need to keep track of in their short-term, working memory.} we propose two-way natural language dialogue with robots as the most near-term, efficient way of communicating with robots 
that does not add to the existing cognitive load on the operator. However, our research goal requires both that we  determine how people would naturally talk to robots (we make no assumptions that this is the same as talking with other people, as prior work has revealed differences \citep{mavridis2015review}), and that we devise methods for robots to understand and produce such language.  
Thus, the focus of this research is the collection and annotation of multi-modal streams of human-robot dialogue to make explicit the patterns of how humans interact with robots and leverage different modalities of information, so as both to establish \textit{common ground} for cooperation and to overcome miscommunications and dynamic changes in the environment (or simply the operator's understanding of 
the environment) that necessitate re-planning. \new{The dialogue annotated here is \textit{multi-floor} dialogue involving multiple conversational participants who share the same high-level dialogue purpose, but there is a unique participant structure and turn-taking expectations within an individual \textit{conversational floor}. Our novel annotation schema over multi-floor dialogue facilitates the development of dialogue systems in disaster-relief and other tasks requiring distributed decision-making and action across heterogeneous teams of people and robots.}

To act as a conversational and cooperative partner in complex, physically-situated tasks like disaster relief, robots must have access to several dimensions of meaning: propositional meaning within an utterance or instruction, the conversational intent of an utterance with respect to those that come before and after, and the interpretation of the utterance with respect to the 
physical environment. We postulate that a robot or, more generally, an agent must have access to each of these 
dimensions of meaning in order to establish and maintain \textit{common ground}---shared mutual knowledge and assumptions \citep{clark81} of what has been said and understood between conversational partners \citep{clark86,clark89}. \new{Common ground can be more challenging to establish and maintain when there are differences in what can be perceived (e.g., when a robot operates at a distance in a bandwidth-limited context without streaming video), as would arise when communications infrastructure are downed in a disaster, unavailable in a remote environment, or disrupted in a contested military environment.}

\begin{figure}[h!]
\begin{itemize}[leftmargin=1.6cm]
\label{excerpt1}
\item[{\bf CMD$_{1}$}] \new{\textit{``robot turn forty five degrees to the left"}}
\item[{\bf CMD$_{2}$}] \new{\textit{``and take a photo"}}
\item[{\bf Robot$_{3}$}] \new{\textit{``executing..."}}
\item[{\bf Robot$_{4}$}] \new{\textit{``sent"}}
\noindent \begin{flushright}(\new{Excerpt 1})\end{flushright}
\end{itemize}
    \centering
    \begin{subfigure}[t]{0.45\textwidth}
        \includegraphics[width=\textwidth]{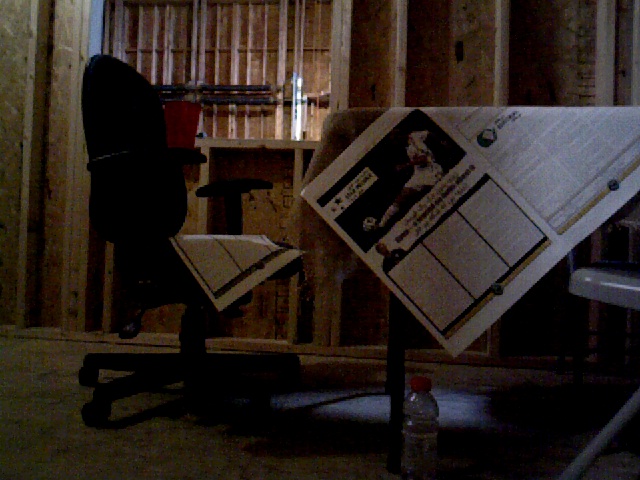}
        \caption{\label{photo-new1}\new{Photo seen by Commander when issuing instruction in Excerpt 1, CMD$_{1}$}}
    \end{subfigure}
    ~
    \begin{subfigure}[t]{0.45\textwidth}
    \centering
        \includegraphics[width=\textwidth]{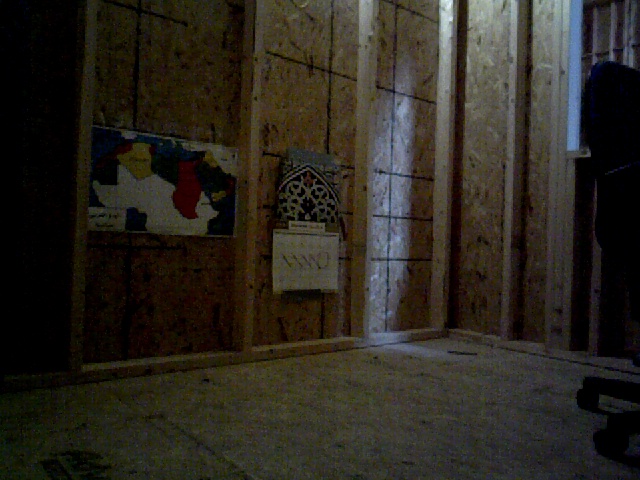}
        \caption{\label{photo-new2}\new{Photo seen by Commander after receiving the photo in Excerpt 1, Robot$_{4}$}}
    \end{subfigure}
    \caption{Images taken during Excerpt \hyperref[excerpt1]{1}, from SCOUT corpus}
    \label{photo-example-new}
\end{figure}

Consider Excerpt~\hyperref[excerpt1]{1} 
from the Situated Corpus of Understanding Transactions (SCOUT), which consists of human-robot dialogues between a human Commander (CMD) and a Wizard-of-Oz robot \citep{lukin2024scout}. 

\new{Over the course of an exploratory search exercise with limited bandwidth communication, the human Commander gives verbal instructions such as CMD$_{1-2}$ to a remotely located robot that responded through text messages in Robot$_{3-4}$. The robot cannot stream its video feed, but it can send photos upon request, as well as a low-resolution 2-dimensional map and position data from LIDAR (Light Detection and Ranging sensor). In Excerpt~\hyperref[excerpt1]{1}, the Commander views the most recent snapshot sent from the robot, Figure~\ref{photo-new1}---which had been taken as a result of the Commander's prior request for a photo---and uses it as a window of shared common ground to issue instruction CMD$_{1}$. The robot successfully interprets and executes CMD$_{1}$ and takes and sends a new photo from the new location per CMD$_{2}$, resulting in Figure~\ref{photo-new2}.}

\begin{figure}[t!]
\begin{itemize}[leftmargin=1.6cm]
\label{excerpt2}
\item[{\bf CMD$_{5}$}] \textit{``robot continue down the hallway directly in front of you underneath the overhead light"} 
\item[{\bf Robot$_{6}$}] \textit{``I don’t see an overhead light in my current position. Would you like me to send a photo?"}
\item[{\bf CMD$_{7}$}] \textit{``robot send a photo"}
\item[{\bf Robot$_{8}$}] \textit{``sent"}
\item[{\bf CMD$_{9}$}] \textit{``robot continue moving forward to the right of the red bucket"} 
\noindent \begin{flushright}(\new{Excerpt 2})\end{flushright}
\end{itemize}
    \centering
    \begin{subfigure}[t]{0.45\textwidth}
        \includegraphics[width=\textwidth]{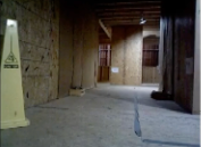}
        \caption{\label{fig:photo-example1}Photo seen by Commander when issuing instruction in \new{Excerpt 2}, CMD$_{5}$}
    \end{subfigure}
    ~
    \begin{subfigure}[t]{0.45\textwidth}
    \centering
        \includegraphics[width=\textwidth]{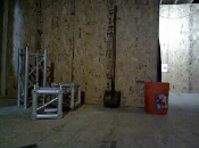}
        \caption{\label{fig:photo-example2}Photo seen by Commander when issuing instruction in \new{Excerpt 2}, CMD$_{9}$}
    \end{subfigure}
    \caption{Images taken during \new{Excerpt \hyperref[excerpt2]{2}}, from SCOUT corpus}
    \label{photo-example}
\end{figure}

\new{However, ambiguity can arise as a result of misunderstanding, multiple referents, or a loss of visual common ground. Consider a different dialogue in Excerpt~\hyperref[excerpt2]{2} and the photos sent (Figure~\ref{photo-example}). In this case, the robot has moved since the photo in Figure~\ref{fig:photo-example1} was taken, so a need arises to reestablish common ground, which can be accomplished with a dialogue interaction.} In order to interpret and act upon the initial instruction CMD$_{5}$, a robot must be able to interpret the propositional content of the instruction---what does \textit{continue} signify in this context (a movement event), and what are the parameters or roles of that event? However, note that an awareness of the conversational intent of the utterance is also required---here, the utterance is a command (the action is given in imperative form, \textit{continue,} as opposed to the conjugated \textit{continues}) that ideally should be 
responded to with an execution of the command\new{, or feedback as to whether and when it will be done, or why it can't be done and how to achieve the high-level intent}.

This awareness of the conversational intent of each utterance is equally important in understanding the follow-up observation, \textit{I don't see an overhead light in my current position} and the subsequent offer, \textit{Would you like me to send a photo?}.  When taken together, these responses make clear why the robot is not able to respond to the initial command by executing it, \new{as desired}. 
The robot attempts to repair the problem and re-establish common ground with a picture of what it sees with its camera from its current position. 

Finally, 
one can see that beyond conversational intent, there is the requirement for \textit{symbol grounding}---\new{defining the meaning of natural language concepts (symbols) in terms of meaningful input to the robot, as opposed to defining those symbols as other words or symbols \citep{HARNAD1990335}. Given the modalities available to the robot, we are specifically interested in reference resolution (picking out the referent of a referring expression) and visual grounding (picking out which part of an image would be a \textit{hallway} based on visual and LIDAR sensory input).  A single instruction can draw upon symbol grounding generally (e.g., what arrangement of sensory inputs correspond to the spatial relation \textit{left}), reference resolution (e.g., which door is being referred to in the environment), and visual grounding (e.g., which area of an image corresponds to \textit{The door on the left in the last image you sent}). Thus, for simplicity, we will refer to all of these as ``grounding.''\footnote{Note that we discuss two related, but distinct processes of \textit{grounding}. The process of establishing \textit{common ground} is arriving at shared mutual knowledge and assumptions between interlocutors. \new{The processes described above, including symbol grounding, visual grounding, and reference resolution, occur within an individual, trying to reconcile perceptual stimuli with each other and with cognitive representations and actions. These two types of grounding often co-occur, as the internal processes may be necessary to understand and engage in dialogue to build common ground, and dialogue contributions present symbolic inputs that must be reconciled with visual inputs and concepts.}}
Both the interpretation and execution of the instruction CMD$_{5}$ are precluded without the ability to successfully ground the words \textit{hallway} and \textit{overhead light}.}
Furthermore, the robot must ground the expression \textit{continue} to one of its own executable behaviors, here forward locomotion. 

We note also that a human-robot dialogue
system that is unable to dynamically track conversational intent across turns in a two-way dialogue would not be able to 
overcome the loss of common ground that has apparently taken place at the utterance of CMD$_{5}$---the 
overhead light is not visible to both parties. Recognizing this loss requires inference over the environment and an attempt to (symbol) ground the instruction before the robot attempts to execute the instruction; the alternative would be a situation in which the grounding and planning fail, but there is no recourse to dialogue to overcome the problem. For this reason, multi-modal two-way dialogue is needed to pinpoint the problematic portion of CMD$_{5}$ and provide the interlocuter with the resources needed to re-establish common ground, which is achieved through the updated photo \new{in Figure~\ref{fig:photo-example2},}
and the repaired instruction in CMD$_{9}$.

\new{To engage in such purposeful, two-way dialogue, a robot has to be able to understand and reason about several aspects of utterance meaning and its relation to the context in which it was uttered.  First, it must represent the illocutionary force of an utterance (is it an action to be performed, or information to be integrated, or something else to be responded to) and the content (what exactly is to be done), and how these relate to other possible meanings. There must also be an understanding of how each utterance relates to others by the same and other speakers---is it about the same thing as previous messages? Is it feedback about how a previous message was understood, or confirmation that actions have been executed, or clarification? Relations help build common ground or make clear that it is not present. Finally, in order to act appropriately, there must be a correspondence of meaning across modalities, so that one can draw correspondences between verbal descriptions and visual depiction of objects and the space they are in. Images and maps are often necessary to fully understand what would otherwise be abstract instructions sufficiently to execute them as intended. For example, does an expression like \textit{the door on the left} refer to a uniquely identifiable door, or ambiguously to multiple candidates on the left, or is there no appropriate door in that position?}

\begin{figure}
    \centering
    \includegraphics[width=4.8in]{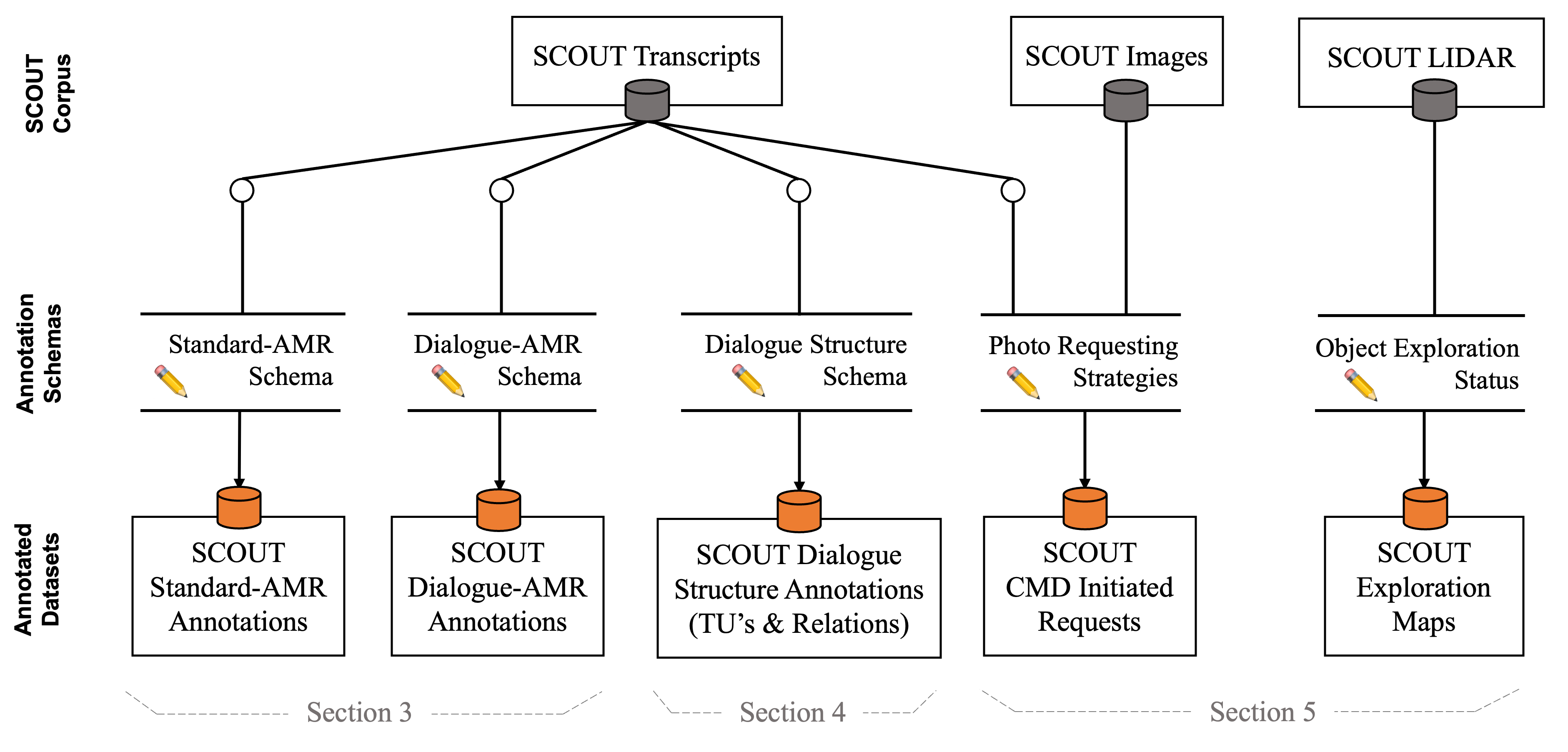}
    \caption{Annotated corpora and paper roadmap}
    \label{fig:coropora-fig}
\end{figure}

In order to make each of these dimensions of meaning accessible and enable common ground, we annotate our SCOUT corpus with four types of linguistic annotation  
and two types of visual annotation as depicted in Figure~\ref{fig:coropora-fig}.  \new{To assess each interlocutor's intent and meaning within a single utterance, we annotate SCOUT with both (1) Abstract Meaning Representation (AMR) \citep{banarescu2013abstract} and (2) Dialogue-AMR \citep{bonial2019abstract}.  To reveal patterns of the Dialogue Structure \citep{traum2018dialogue} across utterances in multi-floor dialogue, we annotate (3) Transactional Units (TUs) \citep{carletta-etal:1996} and (4) the Relations between utterances within a TU; together, these Dialogue Structure annotations pave the way for robot engagement in multi-party cooperative tasks, where the parties may be heterogeneous teams of humans and robots.} To model the way in which interlocutors leverage different modalities, specifically visual information (images, LIDAR), in situated dialogue, we interleave into the dialogue transcripts, the photos of the environment requested by Commanders and the strategy or motivation for requesting that image. Finally, we annotate LIDAR maps---a persistent and shared resource between the human and robot--and construct an Exploration Map of the environment which serves to inventory the status of which objects of interest have been scanned as located on the map. \new{For annotations of both utterance meaning and dialogue structure we consider both prior well-developed ontologies and annotation schemas as well as the special requirements of our domain, as revealed by language usage in the corpus. Our annotations are thus a combination of existing schemas (e.g., AMR, Transactions), extensions to include additional relevant phenomena (Dialogue-AMR), and new schemas influenced by prior work (dialogue relations, which leverage some aspects of both the ISO schemas for Dialogue acts and discourse relations).}

In the section to follow, we describe the human-robot domain of search and navigation (Section~\ref{sec:space}). The subsequent sections of this article then describe the development of our annotation schemas on the collected multi-modal data streams: 
\begin{itemize}
    \item \new{{\it Standard-AMR} and {\it Dialogue-AMR Annotations} (Section~\ref{sec:amr}): an utterance-level representation of what the speaker is trying to do with the utterance in the conversational context;  distills the content of that utterance into robot behaviors and their parameters. 
     \item 
    {\it Dialogue Structure Annotations} (Section~\ref{sec:dialogue-structure}): a dialogue-level representation of which utterances are related to one another (TUs) and how (Relations) within multi-floor dialogue.}
    \item {\it Visual Context Annotations} (Section~\ref{sec:multi-modal}): photo request strategy annotation of image requests and annotated Exploration Maps from LIDAR.
\end{itemize}
In each annotation section, we close by revisiting \new{Excerpt \hyperref[excerpt2]{2}} to illustrate concretely 
the application of each annotation schema to the example, and thereby provide insights into the 
strengths and shortcomings of each standalone annotation. \new{Although each annotation schema has been independently documented (these papers will be referenced throughout each annotation section), this paper constitutes the first unified description of all levels of annotation, which provides an opportunity for postulating the ways in which the annotations are complementary, as well as considering new insights into where gaps remain in} facilitating a contextualized interpretation 
of language in two-way dialogue and allow cross-modal association of information in maintaining common ground. 
We discuss how these annotations have, independently or in a combined fashion, been put to use, and how we envision enabling the
interpretation of language 
that accesses the different dimensions of common ground, meaning, and architectural components of a dialogue system and robot sensors dynamically (Section~\ref{sec:use-case}). The paper ends with an overview of related work, our conclusions, and brief notes on future work.

\section{Problem Space: Human-Robot Dialogue for Search \& Navigation}
\label{sec:space}

\new{Robots are being increasingly used in disaster relief and robot rescue as robotics technology improves \citep{habibian2021design,edlinger2022intuitive,kanazawa12023considerations}; however, there remain many challenges for supporting natural language communication between a human and their remotely located robot teammate. The first challenge is one of practicality: reliable internet connectivity may not be available to stream high-fidelity video from a robot's sensors due to the nature of the incident (e.g., search and rescue in heavily forested areas, or in cities with broken power lines following a storm or wildfire). In the absence of video, alternative paradigms for sharing visual common ground are necessary in order for the human and robot to communicate in near real-time, such as sharing lower-quality mapping data that shows obstacles with no fidelity of objects, or sending lower-quality photos when the human teammate requests to see an update of how the situation around the robot has changed.}

\new{Beyond these technical challenges, it remains unknown as to \textit{how} a human would want to best utilize a robot teammate in these scenarios to elicit and establish common ground. With no existing human-robot or human-human dialogue corpora in these disaster domains, we cannot design a robot with the most appropriate communication strategies because they are unknown. To this end, we used a Wizard-of-Oz (Woz) experimental paradigm in this problem space to serve multiple purposes: 1) a process by which to collect data from human participants in scenarios involving dialogue and remote instruction-giving, 2) a basis for refining the robot's responses to the human, and 3) a corpus of training data for automating components of the robot to achieve our goal of autonomous communication in a search and reconnaissance scenario. Despite the limitations inherent in the scenario itself, e.g., no streaming of video, we interpret the language of the human as meaningful as they adapt to overcome these limitations and complete the task.}

Here we describe the search and navigation task in the SCOUT dataset (the Situated Corpus On Understanding Transactions) \citep{lukin2024scout}, \new{which was designed to meet these problem space criteria above,} and the features of the human-robot dialogue and interaction that are key to our annotation of aspects contributing to common ground. 
In SCOUT, a human operator, or Commander, instructs what they believe to be an autonomous robot in a remote location through a series of search and navigation tasks including counting and finding doorways, shovels, and shoes in an abandoned house and detecting evidence that the location has been recently occupied. The Commander speaks to the robot in natural language while sitting at a workstation with three sources of information: a chat stream of text replies from the robot; a 2D, birds-eye view map of the environment from the robot's onboard LIDAR (LIght Detection and Ranging) laser scanner; and an image taken at the Commander's request from a static, front-facing RGB camera on the robot. The Commander is not given any instructions about how to speak with the robot (other than how to use the push-to-talk microphone provided). 

\new{The robot used is a Clearpath Jackal Unmanned Ground Vehicle. Its capabilities are limited to driving forward or backward, and rotating to the left or right. The robot does not have arms or manipulators, and cannot hear or emit sound. The Commander is shown a picture of the robot, and can discover its capabilities by trying out different instructions and observing the robot. In cases where the robot is asked to do something it cannot, it responds to the Commander as such.}

\begin{figure}[t!]
\centering
\includegraphics[width=4.5in]{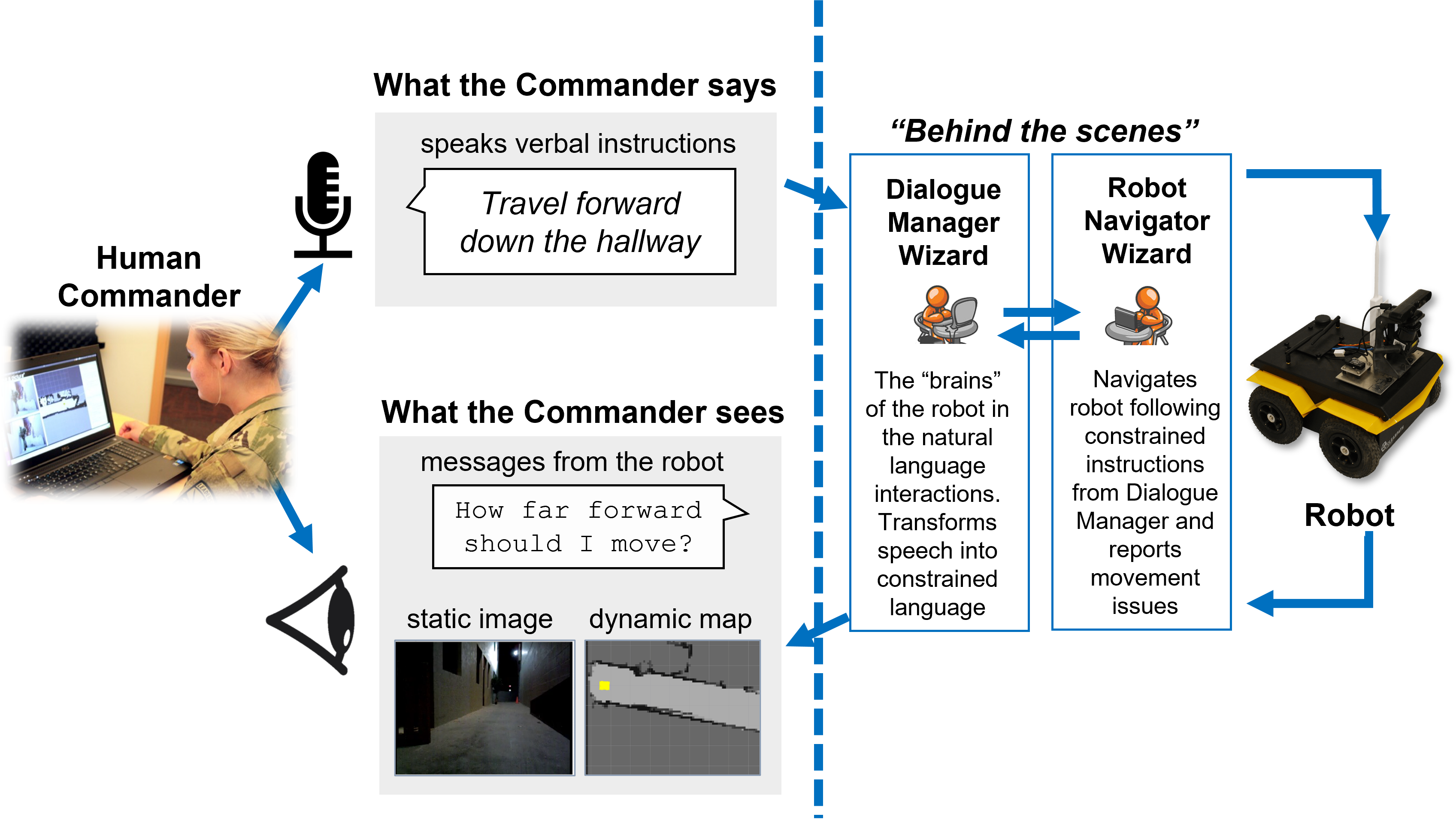}
\caption{The Commander issues verbal
instructions to the robot, whose capabilities are
performed by two wizards standing in for the respective, separate
abilities of dialogue management and robot navigation. Original figure from~\cite{lukin2024scout}.}
\label{expt-layout}
\end{figure}

Behind the scenes, experimenters control the robot's dialogue processing and robot navigation capabilities to differing extents, depending upon the experimental phase. This methodology was inspired by the success of research using a Data-driven Wizard-of-Oz methodology to observe how humans would chat with what they believed to be an autonomous virtual avatar \citep{devault2014simsensei}. In the same way, Commanders in SCOUT spoke to a robot they believed to be autonomous, when it was in fact controlled by multiple Wizards. The Dialogue Manager (DM) wizard stood in for the understanding and dialogue management components of our system by interpreting the Commander's instructions, selecting responses, and passing the Commander's intent along to another wizard, namely the Robot Navigator (RN) wizard, who stood in for the planning and motor execution components of our system by joysticking the robot to complete the instruction. The DM-Wizard interacts directly with the Commander, while the RN-Wizard only interacts with the DM-Wizard (never the Commander). \new{This mimics the nature of what a final, all-automated version of the system would look like, in which an automated dialogue management component would communicate with the human user and pass parsed instructions to the physical motion component, while the physical motion component would not speak directly with the user. }The DM-Wizard had access to the last image the Commander requested, a live video stream from the robot's camera, and the same 2D LIDAR map the Commander could see. The RN-Wizard had access to the live video stream and a 3D version of the LIDAR map. For a depiction of this setup, see Figure~\ref{expt-layout}. 

\new{Efforts to build} SCOUT involved three experiments which utilized this multi-WoZ setup. In the fourth and final experiment of our project developing SCOUT, the DM-Wizard was replaced with an automated dialogue manager.
Each Commander (a participant in one of our experiments) completed three trials: one considered training and two \new{involving larger main environments}, in which the objectives and target objects of interest varied.

\begin{table*}[h]
\centering
\begin{small}
\caption{Multi-Floor Dialogue of \new{Excerpt \hyperref[excerpt2]{2}} depicting the conversational floors and ``behind the scenes'' Wizard dialogue.}
\label{tab:excerpt1-table}
\begin{tabular}{llp{1.2in}p{1.2in}p{0.6in}p{0.6in}}

\toprule
& \multicolumn{2}{c}{Left Floor}              
& \multicolumn{2}{c}{Right Floor} \\

\cmidrule(r{1em}l{0.75em}){2-3}
\cmidrule(rl{0.75em}){4-5}

\# & Time (s) & \textbf{CMD} & \textbf{DM$\rightarrow$CMD} & \textbf{DM$\rightarrow$RN} & \textbf{RN} \\ 
\midrule
  
1 & 0 & robot continue down the hallway directly in front of you underneath the overhead light& & & \\ \hline
\rowcolor{gray} 2 & 45.75 & & I don't see an overhead light in my current position.  Would you like me to send a photo? & &  \\ \hline
3 & 47.17 & robot send a photo & & &  \\ \hline
\rowcolor{gray} 4 & 52.65 & & & photo & \\ \hline
5 & 53.47 & & & & image sent \\  \hline
\rowcolor{gray} 6 & 54.31 & robot continue moving forward to the right of the red bucket& & & \\ \hline
\end{tabular}
\end{small}	
\end{table*}

Table~\ref{tab:excerpt1-table} shows \new{Excerpt \hyperref[excerpt2]{2}} again, this time transcribed in a multi-floor depiction between all actual (Commander (CMD), DM-Wizard (DM), RN-Wizard (RN)) interlocutors. The dialogue from the ``Robot'' is in actuality the DM speaking to the Commander, and takes place on the {\it left} conversational floor between the Commander and DM. On the {\it right} conversational floor, the DM speaks to the RN to translate instructions ``behind the scenes'' (in utterance \#4, the command is to the RN to take a photo, and in utterance \#5, the RN gives the DM an acknowledgement). The Commander and RN never speak directly to or hear each other; instead, the DM acts as an intermediary passing communication between the Commander and the RN. Again, we emphasize that this multi-floor setup is not merely an artifact of the data collection---it mimics how the physical motion component of an autonomous system would not ``speak'' directly to a human user but rather would receive its parsed instructions from a separate module that handled user interface (here, natural language) interactions. Each separate module is represented by a separate wizard, communicating on separate floors as would occur in a fully automated system. 

Timestamps are provided in Table~\ref{tab:excerpt1-table} reflecting the time passed in seconds from utterance \#1. There is a significant passage of 45s between the Commander's initial instruction and the DM's response (\#2), reflecting the DM's challenge in attempting to understand and provide an adequate response to the ambiguous instruction. The remainder of the dialogue proceeds more quickly, taking only several seconds between utterances as they are passed and responded to across the conversational floors.

The task of the Commander is challenging on many levels. The dialogues in SCOUT are approximately 20-minutes long and on average contain 320 utterances, allowing for an extensive dialogue history that becomes a cognitive burden for the Commander to remember what they have seen, done, and asked for. Because Commanders were not given instructions on how to speak to the robot, their instructions are rich in vocabulary and structure as they first evolved and decided on a strategy for conveying instructions effectively with an extremely unfamiliar conversational partner, and second evolved and decided on a strategy for how one might go about navigating the robot through the unfamiliar environment to attempt to address their search tasks. 

The dialogues are situated, with the multi-modal information streams available in the SCOUT transcripts with the images requested by the Commander throughout the dialogue. The Commander could only view the last image they requested, which simultaneously gave them a window into the environment at any point in time, while limiting their current field of view. The images were the only way certain items could be found and counted to fulfill the experimental task. \new{This choice was purposeful and meant to induce greater reliance on the dialogue system for accomplishing the search tasks in the constrained bandwidth conditions.}

The SCOUT \new{corpus} contains 89,056 utterances and 310,095 words from 278 dialogues averaging 320 utterances. The dialogues in SCOUT are aligned with the multi-modal data streams available during the experiments: 5,785 images and a subset of 30 maps. \new{Additional statistics relating to the corpus statistics can be found in \cite{lukin2024scout}.}

\section{\new{Dialogue-AMR Annotations}}
\label{sec:amr}

\begin{figure*}[!ht]
\centering
\includegraphics[width=4.5in]{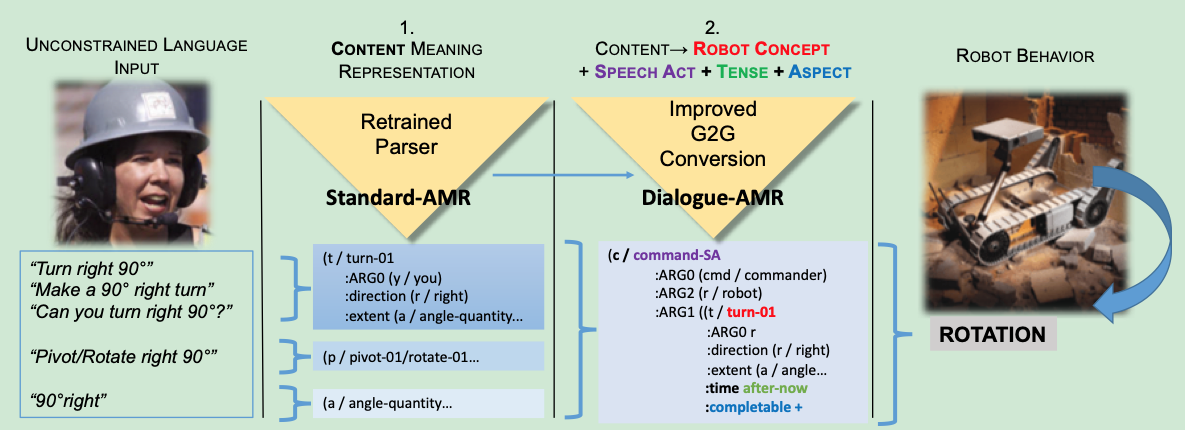}
\caption{Verbal instructions are parsed into Standard-AMR using automated parsers, then converted into Dialogue-AMR, and if executable, mapped to a robot behavior.} 
\label{fig:pipeline}
\vspace{-0.1in}
\end{figure*}

Here, we describe a schema that employs and enriches Abstract Meaning Representation (AMR) \citep{banarescu2013abstract} to support Natural Language Understanding (NLU) in human-robot dialogue systems
(see Figure~\ref{fig:pipeline}).
We hypothesize that a semantic
representation \new{of participant instructions} is needed in the process of mapping natural language instructions to an autonomous system's set of executable behaviors and to the real-world objects mentioned in those
behaviors (\new{a process we shall refer to as ``grounding'')},
 we developed, ``Dialogue-AMR,''  
an augmentation of AMR specifically for human-robot dialogue.

AMR, which we will refer to as ``Standard-AMR'' to clarify the distinction from Dialogue-AMR, is a formalism for sentence semantics that abstracts away many syntactic idiosyncrasies and represents sentences with rooted, directed, acyclic graphs (Figure~\ref{fig:template-intro}a shows the PENMAN notation of a Standard-AMR graph). Although Standard-AMR provides a suitable level of abstraction for representing the content of sentences, it lacks several aspects of meaning crucial to our domain. 
Dialogue-AMR 
adds to Standard-AMR information on the speaker's intent, or ``illocutionary force,'' as well as tense and aspect information.
Additionally, 
Dialogue-AMR translates a variety of different expressions for the same underlying behaviors (e.g., \textit{go, move, drive}), to a single, uniform 
predicate (e.g., \texttt{go-02} in Figure~\ref{fig:template-intro}b). Thus, it allows the system to distill the action primitives and their parameters from 
unconstrained natural language and enables the subsequent grounding of this information in the robot's 
capabilities and environment. 

\begin{figure}[!ht]\begin{small}
\begin{verbatim}

(a) (d / drive-01 :mode imperative
        :ARG0 (y / you)
        :destination (d2 / door))
(b) (c / command-SA
        :ARG0 (c2 / commander)
        :ARG2 (r / robot)
        :ARG1 (g / go-02 :completable +
           :ARG0 r
           :ARG3 (h / here)
           :ARG4 (d/ door)
           :time (a2 / after
                :op1 (n / now))))
\end{verbatim}
\end{small}
\caption{The utterance \textit{Drive to the door} represented in (a)~Standard-AMR form and (b)~Dialogue-AMR form. }
\label{fig:template-intro}
\end{figure}

\subsection{Annotation Description}
To develop augmentation of Standard-AMR that addresses the requirements 
in human-robot dialogue, we iteratively refine 
an inventory of speech acts (Section~\ref{ssec:SpeechActInventory}) and introduce tense and aspect representations 
not included in Standard-AMR (Section~\ref{subsec:tenseaspect}). 
These additional elements of meaning are brought together in our annotation schema for Dialogue-AMR, in which the propositional content is also 
normalized by replacing a variety of lexical items in the input language (e.g., \textit{turn, pivot, rotate}) with an assigned semantic relation (e.g., \amr{turn-01}) that in turn maps to a single robot concept (e.g., \textsc{rotation}) corresponding to one of the robot's executable behaviors (Section~\ref{ssec:annotationschema}). 

\subsubsection{Speech Act Inventory} 
\label{ssec:SpeechActInventory}

Speech acts are introduced as the root of Dialogue-AMR representations, such as \texttt{Command-SA} in 
Figure~\ref{fig:template-intro}b, where the suffix -SA is an abbreviation for ``Speech Act.'' 
Adding information on the speech act captures the illocutionary force, or what the speaker is trying to do 
in the conversational context with their individual utterance. 
For example, a request for information and a request for action serve distinct dialogue functions. Similarly, a promise regarding a future action and an assertion about a past action update the conversational context in very different ways.

\new{While extensive and reliable annotation schemas for dialogue acts already exist (e.g., \cite{bunt2012iso}), these are not always in perfect alignment with our data. There are many distinctions in prior schemas that are not seen in our data, while other important distinctions regarding relationships between utterances, perception and action are not elaborated. To simplify some of these issues, we adopted the following approach. First, following Damsl guidelines \cite{Damsl,core1997coding}, we distinguish between forward-looking function from backward-looking function. Forward looking function concerns how the current utterance constrains the
future beliefs and actions, and is expressed as the speech acts in Dialogue-AMR described in this section. In contrast, backward-looking function concerns relations to prior utterances, (e.g. feedback and various types of responses), and is discussed in section~\ref{sec:dialogue-structure}. Second, each of these schemas are developed with reference to distinctions from the literature but with a focus on issues that arise in the data.}

\begin{figure*}[t]
\begin{center}
\includegraphics[width=4.85in]{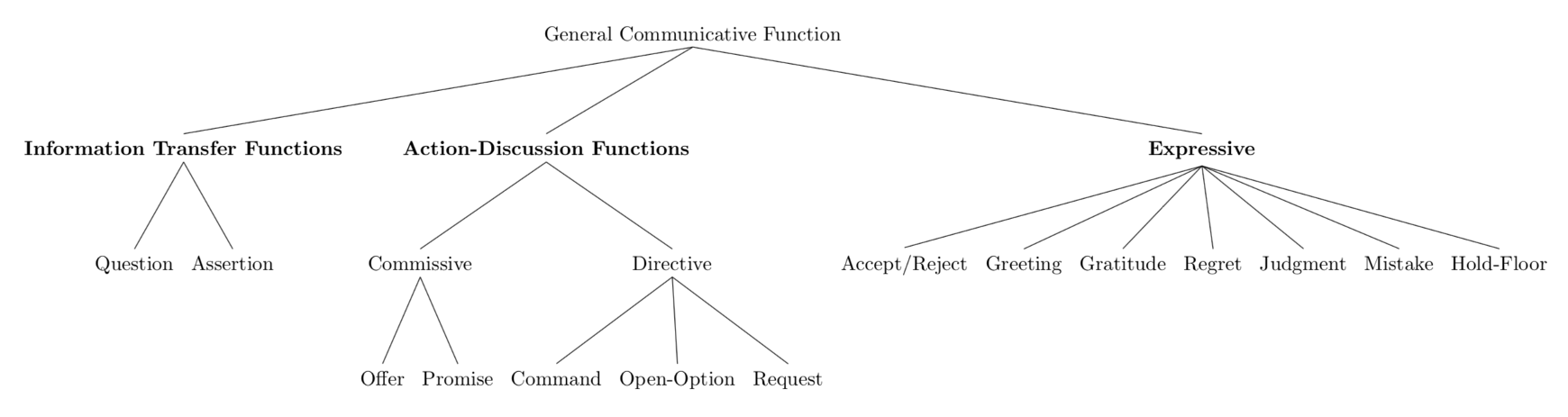} 
\caption{Dialogue-AMR Speech Act Taxonomy}
\label{fig:SATaxonomy}
\end{center}
\end{figure*}

The Dialogue-AMR inventory of speech acts incorporates much of the higher-level categorization and labeling of speech acts outlined by \cite{searle1969speech}, including the basic categories of Assertions (termed ``representatives'' by Searle), 
Commissives, Directives, and Expressives. 
Additionally, based on \cite{bunt2012iso}, we introduce an early distinction in classifying our speech acts between Information Transfer Functions and Action-Discussion Functions (see Figure~\ref{fig:SATaxonomy}). 

In terms of dialogue function, these higher-level categories allow us to monitor the status of distinct dialogue contexts. For Information Transfer types, we can monitor the quantity and quality of general-purpose information exchanged in the dialogue that is relevant to the larger task at hand. For example, \textit{Robot, do you speak any foreign languages?} may not directly impact a current task, but it introduces information into the dialogue that may be useful at a later point. For Action-Discussion types, we can assess the status of individual tasks as the dialogue progresses. For example, (\textit{Moving to the wall}) and (\textit{I moved to the wall}) convey two points on a timeline related to current task completion. For Expressive types, we can model the changing relationship between interlocutors---for example, how utterances of gratitude, acceptance or rejection, and admission of mistakes impact the level of trust between the two interlocutors. 

Beyond these higher-level categories, we iteratively refined the speech act categories needed for our domain based upon rounds of surveying and annotating our data.   These iterations began with the annotation of ``dialogue moves'' over participant instructions only \citep{marge2017exploring} and evolved with varying numbers and types of speech acts \citep{bonial2019augmenting} to the inventory set forth here.

Table~\ref{tab:RobotLexicon} lists the relation integrated into the Dialogue-AMR to represent the speech acts. 
\new{We propose a new set of speech act relations, which were not included in Standard-AMR, denoted by the ending \amr{-SA} in lieu of the numbered endings currently included in Standard-AMR relations}. Although we originally explored adopting existing AMR relations that best fit with each speech act (e.g., \amr{Question-01, 
Command-02}) \citep{bonial2019augmenting}, we subsequently opted to introduce new relations so that the Dialogue-AMR is clear in what portion represents propositional content and what portion represents the illocutionary force.\footnote{The corpus release 
includes a mapping allowing for conversion of SA relations into 
existing Standard-AMR numbered relations.}
Additionally, we found that existing AMR relations were inconsistent in the 
argument structure representing the speaker, addressee, and content of 
the speech act. For example, while \amr{Command-02} represents the addressee or impelled agent as \amr{Arg1} and the impelled action as \amr{Arg2}, we found \amr{Assert-02} instead represents the addressee as \amr{Arg2} and 
the content of the assertion as \amr{Arg1}. 
Our roles in our speech acts
maintain the following consistent argument structure (as seen in Figure~\ref{fig:template-intro}b): \\ \\
\hspace*{1em}\amr{Arg0: Speaker}\\
\hspace*{1em}\amr{Arg1: Content}\\
\hspace*{1em}\amr{Arg2: Addressee}\\

\noindent The roles of \amr{Arg0} and \amr{Arg2} correspond consistently to Speaker 
and Addressee, respectively; the semantics of the \amr{Arg1} shifts depending 
upon the particular speech act. For example, the \amr{Arg1-content} of 
\amr{Command-SA} is an action, whereas the \amr{Arg1-content} of 
\amr{Regret-SA} is the stimulus of the mental state, or the thing 
regretted. 

\subsubsection{Tense and Aspect in Dialogue-AMR}\label{subsec:tenseaspect}

There are patterned interactions between tense and aspect and illocutionary force that are critical for conveying the robot's current status in our domain. These include the distinctions between: a promise to carry out an instruction in the future, a declarative statement that the instruction is being carried out currently, and an acknowledgment that it was carried out in the past.
Standard-AMR lacks information that specifies \textit{when} an action occurs relative to speech time, and whether or not this action is completed (if a past event), or is able to be completed (if a future event). 
For example, Standard-AMR represents the common feedback utterances (\textit{I will move forward 10 feet}), (\textit{I am moving\dots}), and (\textit{I moved\dots}) with one identical graph (see Figure~\ref{fig:identicalamr}). 

\begin{figure}[!ht]\begin{small}
\begin{verbatim}
(m / move-01
      :ARG0 (i / i)
      :direction (f / forward)
      :extent (d / distance-quantity 
            :quant 10
            :unit (f2 / foot)))
\end{verbatim}
\end{small}\caption{Because Standard-AMR lacks tense and aspect representation, the phrases \textit{I will move / I am moving / I moved... forward 10 feet} are represented identically, as shown here.}
\label{fig:identicalamr}
\end{figure}

We incorporate tense and aspect information into Dialogue-AMR by adopting the  annotation schema  
of \cite{donatelli2018annotation}, who propose a four-way division of temporal annotation and four multi-valued categories for aspectual annotation that fit seamlessly into existing AMR annotation practice. We reduced Donatelli et al.'s proposed temporal categories to three,\footnote{Eliminating the \amr{up-to} temporal relationship.} in order to capture temporal relations before, during, and after the speech time. In addition to the aspectual categories they proposed, we added the category \amr{:completable~+/-} to signal whether or not a hypothetical event has an end-goal that is executable for the robot (see \cite{donatelli2019tense} for a sketch of this aspectual category). 
Our annotation categories for tense and aspect can be seen in Figure~\ref{fig:tenseaspectlabels}.\footnote{The \amr{:habitual} aspectual category is absent from the current annotated data. However, we maintain it as a possible category in anticipation of future work and the potential to refer to habitual robot actions. \new{We acknowledge that certain pairings of aspectual features happen frequently and may seem redundant (e.g., \amr{complete -} and \amr{ongoing +}). For clarity and completeness, we mark all telic events (those with a clear ending point) as \amr{complete +/-} as distinct from states, which are marked as \amr{:stable +/-}, and habitual events (\amr{:habitual +/-}), which may also interact with \amr{:ongoing +/-}. For full details, see \citet{bonial2023dialogue}}.}
\begin{figure}[h]
   \centering\small 
   \begin{minipage}[t]{0.55\columnwidth}
     \textsc{temporal annotation} \\[1ex]
     \begin{tabular}{|l|} \hline
       \Time{\amr{:time}} \\
       \hspace{2mm}1. \Time{\amr{(b / before}} \\
       \hspace{10mm}\Time{\amr{:op1 (n / now))}} \\
       \hspace{2mm}2. \Time{\amr{(n / now)}} \\
       \hspace{2mm}3. \Time{\amr{(a / after}} \\
       \hspace{10mm} \Time{\amr{:op1 (n / now))}} \\\hline
     \end{tabular}
   \end{minipage}%
   \begin{minipage}[t]{0.45\columnwidth}
     \textsc{aspectual annotation} \\[1ex]
     \begin{tabular}{|l|} \hline              
       \Aspect{\amr{:stable +/-}} \\                           
       \Aspect{\amr{:ongoing +/-}} \\                          
       \Aspect{\amr{:complete +/-}} \\                         
       \Aspect{\amr{:habitual +/-}} \\                         
       \Aspect{\amr{:completable +/-}} \\\hline
     \end{tabular}
   \end{minipage}
   \caption{Three categories for temporal annotation and five categories for aspectual annotation are used to augment existing AMR for collaborative dialogue. }
   \label{fig:tenseaspectlabels}
\end{figure}

Notably, this annotation schema is able to capture the distinctions missing in Figure~\ref{fig:identicalamr}. Updated AMRs for utterances that communicate information about a \textsc{movement} event relative to the future, present, and past are shown in Figure~\ref{fig:amrwithtenseaspect}. 

\begin{figure}[!ht]\begin{small}
\begin{verbatim}
1. (m / move-01 :completable +
      :ARG0 (i / i)
      :direction (f / forward)
      :extent (d / distance-quantity 
            :quant 10
            :unit (f2 / foot))
      :time (a / after
            :op1 (n / now)))
            
2. (m / move-01 :ongoing + :complete -
      :ARG0 (i / i)
      :direction (f / forward)
      :extent (d / distance-quantity 
            :quant 10
            :unit (f2 / foot))
      :time (n / now))

3. (m / move-01 :ongoing - :complete +
      :ARG0 (i / i)
      :direction (f / forward)
      :extent (d / distance-quantity 
            :quant 10
            :unit (f2 / foot))
      :time (b / before
            :op1 (n / now)))

\end{verbatim}
\end{small}\caption{Updated AMRs for (1) \textit{I will move...}, (2)\textit{ I am moving...}, and (3) \textit{ I moved...}. Compare with Figure~\ref{fig:identicalamr} for added tense and aspect information.}
\label{fig:amrwithtenseaspect}
\end{figure}
Using the schema presented in Figure~\ref{fig:tenseaspectlabels}, our Dialogue-AMRs allow for locating an event in time and expressing information related to the boundedness of the event (i.e. whether or not the event is a future event with a clear beginning and endpoint, a present event in progress towards an end goal, or a past event that has been completed from start to finish).

\subsubsection{Full Annotation Schema in Dialogue-AMR}
\label{ssec:annotationschema}
Our meaning representation is intended to bridge the gap from totally unconstrained natural language input to the appropriate action specification in the robot's 
limited repertoire, including clarification actions. 
In order to understand an input utterance such that it is actionable, the robot must recognize both the illocutionary force and the propositional content of the utterance. We integrate both these levels of meaning into a 
single Dialogue-AMR representation.  
The Dialogue-AMRs can be thought of as templates or 
skeletal AMRs in which the top anchor node is a specific relation corresponding to an illocutionary force (e.g., \amr{assert-SA}) and its arguments hold the propositional content of the utterance, where the latter consists of a relation (e.g., \amr{turn-01}, \amr{go-02}) corresponding to an action specification from the robot's repertoire of behaviors (e.g., \textsc{Rotation}, \textsc{Movement}). The relation's arguments are filled in the template based on the specifics of the utterance (see Figure~\ref{fig:template}). 
\begin{figure}[!ht]\begin{small}
\begin{verbatim}
(a) (m / move-01 :mode imperative
        :ARG0 (y / you)
        :ARG1 y
        :ARG2 (w / wall))
(b) (c / command-SA 
        :ARG0-\textit{speaker}
        :ARG2-\textit{addressee}
        :ARG1 (g / go-02 :completable +
            :ARG0-\textit{goer} 
            :ARG1-\textit{extent} 
            :ARG3-\textit{start point }
            :ARG4-\textit{end point} 
            \textit{:path}
            \textit{:direction}
            :time (a / after
                :op1 (n / now))))
(c) (c / command-SA
        :ARG0 (c2 / commander)
        :ARG2 (r / robot)
        :ARG1 (g / go-02 :completable +
           :ARG0 r
           :ARG3 (h / here)
           :ARG4 (w / wall)
           :time (a2 / after
                :op1 (n / now))))
\end{verbatim}
\end{small}
\caption{
The utterance \textit{Move to the wall} represented in (a) Standard-AMR form, (b) Dialogue-AMR template form, and (c) as a filled-in Dialogue-AMR.
}
\label{fig:template}
\end{figure}

In our pipeline (refer back to Figure~\ref{fig:pipeline}), we leverage both 
automatically generated Standard-AMR as well as the Dialogue-AMR to tame 
the variation found in unconstrained natural language and map this to the robot's constrained repetoire of behaviors. While the Standard-AMR abstracts away from some idiosyncratic syntactic variation, it largely maintains the lexical items from the input language. The Dialogue-AMR, in contrast, maps 
 several lexical items to one robot concept corresponding to an action specification. This concept is realized 
 in the Dialogue-AMR using 
a particular AMR roleset that is part of what we term the robot's lexicon. 
Table~\ref{table:rotate} illustrates an example of the translation from input language to the robot concept of \textsc{rotation}. 

\begin{table}[!h]
\centering
\begin{small}
\caption{Unconstrained input language is compared with its somewhat 
generalized form in Standard-AMR, and its consistent representation with a single relation in Dialogue-AMR, corresponding to the concept {\sc rotation} within the robot's repertoire of behaviors.}
\begin{tabular}{lll}
\toprule
\multicolumn{1}{c}{\textbf{Input}}                                      & \multicolumn{1}{c}{\textbf{AMR}} & \multicolumn{1}{c}{\textbf{\begin{tabular}[c]{@{}c@{}}Dialogue-\\ AMR\end{tabular}}}  \\ \midrule
\textit{Turn left 90 degrees.} & \multirow{2}{*}{\makebox[0pt][r]{\ensuremath{\left.\rule{0pt}{3ex}\right\}}~}\amr{turn-01}}          & \multirow{5}{*}{\makebox[0pt][r]{\ensuremath{\left.\rule{0pt}{7.5ex}\right\}}~}\amr{turn-01}}            \\
\textit{Make a left turn.}     &                                   &                                                                                                                               \\
\textit{Rotate left.}                                                     & \amr{rotate-01}                         &                                                                                                                                                             \\ 
\textit{90 degrees left.}      & \amr{:angle-quantity...}                &                                                                                                                                                          \\
\textit{Pivot 90 left.}                                                   & \amr{pivot-01}                                                      &                                                                                        \\ \bottomrule
\end{tabular}
\label{table:rotate}
\end{small}
\end{table}

Although we had originally hypothesized that we could use a fixed 
set of templates to cover all allowable combinations between particular speech acts and particular actions \citep{bonial2019augmenting}, we have since found that our schema is more flexible and robust to expanding our domain when we eschew that approach in favor of a limited set of speech acts, which combine with an easily expandable lexicon of robot behaviors. 
This facilitates coverage of all possible combinations of speech act and robot concepts, as opposed to limiting ourselves to templates corresponding only to what we have seen thus far. 
Although not exhaustive as to what could be seen in the language of our domain, a table detailing which robot concepts readily combine with which speech acts is given in the Appendix in Table~\ref{tab:RobotLexicon}.

\subsection{Dialogue-AMR Corpus Summary}

A total of 569 utterances from SCOUT have been annotated with 
both Standard-AMR and Dialogue-AMR. \new{Interannotator agreement (IAA) was measured at several points during the development of Dialogue-AMR. Over a representative sample of the 290 utterances of the SCOUT data, IAA was at 86.6\% \citep{bonial2023dialogue}, as measured by the Smatch metric for comparisong AMR graph similarity \citep{cai2013smatch}. This agreement exceeds reported Standard-AMR IAA \citep{bonial2018abstract}. }
We note that other existing AMR corpora that have been released are largely from written text, 
including Wall Street Journal and Xinhua news sources, as well as web discussion forum 
data.\footnote{https://catalog.ldc.upenn.edu/LDC2017T10} There is a small amount (about 200 instances) of broadcast news conversation corpora
but none centered around natural dialogue. Thus, 
our work is one of the first efforts to use AMR to annotate dialogue.
There are additional releases of Dialogue-AMR as well, including annotation of the Minecraft Dialogue 
Corpus \citep{narayan-chen-etal-2019-collaborative}, which was also created in service of developing interactive agents, thereby providing a rich resource addition and source of comparison for SCOUT \citep{bonial-etal-2021-builder}. 

\begin{table}

\adjustbox{max width=0.8\textwidth}{%
\begin{tabular}{lp{1.5in}p{2.4in}}
\toprule                

\# & {\bf Utterance} & {\bf Dialogue-AMR} \\ 
\midrule
  
1 & CMD: robot continue down the hallway directly in front of you underneath the overhead light & \begin{verbatim}(c / command-00
 :ARG0 (c3 / commander)
 :ARG1 (g / go-01 :completable -
   :ARG0 r
   :path (h / hallway
     :ARG1-of (f / front-03
       :ARG2 r)
     :location (u / underneath
       :op1 (l / light
         :mod (o / overhead))))
     :time (a / after
       :op1 (n2 / now)))
  :ARG2 (r / robot))
\end{verbatim}
\\ \hline
\rowcolor{gray} 2 & DM$\rightarrow$CMD: I don't see an overhead light in my current position. Would you like me to send a photo? & \begin{verbatim}(a / assert-00
 :ARG0 (r / robot-dm)
 :ARG1 (s / see-01 :polarity - :stable -
   :ARG0 r
   :ARG1 (l / light
      :mod (o / overhead))
      :time (n / now))
 :ARG2 (c / commander))
(o / offer-00
 :ARG0 (r / robot-dm)
 :ARG1 (s / send-image-99 :completable +
   :ARG0 r
   :ARG1 (i2 / in-front-of
     :op1 r)
   :ARG2 c
   :time (a / after
     :op1 (n / now)))
:ARG2 (c / commander))
\end{verbatim}
\\ \hline
3 & CMD: robot send a photo & \begin{verbatim}(c / command-00
 :ARG0 (c2 / commander-dm)
 :ARG1 (s / send-image-99 :completable +
   :ARG0 r
   :ARG1 (i2 / in-front-of
     :op1 r)
   :ARG2 c2
   :time (a3 / after
     :op1 (n2 / now)))
 :ARG2 (r / robot-rn))
 \end{verbatim}
\\ \hline
\rowcolor{gray} 4 & DM$\rightarrow$RN: photo & \begin{verbatim}(c / command-00
 :ARG0 (c2 / commander-dm)
 :ARG1 (s / send-image-99 :completable +
   :ARG0 r
   :ARG1 (i2 / in-front-of
     :op1 r)
   :ARG2 c2
   :time (a3 / after
     :op1 (n2 / now)))
 :ARG2 (r / robot-rn))
\end{verbatim}
\\ \hline
\end{tabular}}
\caption{\label{tab:excerpt1-amr} \new{Excerpt \hyperref[excerpt2]{2}} shown with Dialogue-AMR Annotations} 
\end{table}

\subsection{Dialogue-AMR Annotation Summary}
\label{ssec:DAMR-summary}

\new{In Table~\ref{tab:excerpt1-amr}, we 
revisit \new{Excerpt \hyperref[excerpt2]{2} where} Dialogue-AMR provides an explicit representation of the content of the utterances.} The first instruction, for example, is broken down into a primary action (\texttt{go-02}) and its parameter (the path for motion  
\textit{the hallway directly in front of you underneath the overhead light}). Note that this instruction contains a very complex prepositional phrase. There is some ambiguity as to whether the speaker intended for 
that whole expression to be a path description, 
or potentially a destination \textit{underneath 
the overhead light}. Thus, although the 
Dialogue-AMR provides the basic behavior primitives for the robot and the parameters to be grounded 
in the robot's environment, this example 
illustrates the ways in which the language 
of situated dialogue need always be interpreted 
dynamically with respect to the robot's current environment. When 
one views the left, initial photo in Figure~\ref{photo-example}, which the 
commander participant was viewing when the instruction was issued, one can see that the position 
under the overhead light appears to be the end of the hallway, thereby motivating an interpretation of this expression as a destination point. See Section~\ref{ssec:amr} for a discussion of how these annotations have been used to implement language understanding and 
grounding in autonomous systems.

Of course the primary source of miscommunication in this excerpt is the fact that \textit{hallway...underneath the overhead light} 
cannot be grounded at all given the robot's current position, as the overhead light is no longer in view. 
\new{A system requires the ability to model information from the prior context 
to} overcome such a mismatch in the two interlocutors' shared understanding of the environment and regain common ground. \new{The Dialogue Structure annotations we describe in the next Section begin to address this gap.}
\new{We further address how references from the visual modality would be injected into the communication effectively in Section~\ref{sec:multi-modal} and will also return to this discussion in Section~\ref{ssec:amr} in the context of grounding for autonomous systems. In the next section, we widen the aperture of the interaction, moving from a focus on 
meaning and illocutionary force within a single utterance to a focus on cross-utterance relations and structure. }

\newpage

\section{\new{Dialogue Structure Annotations}}
\label{sec:dialogue-structure}

Here, we present our annotation scheme for \textit{meso-level} dialogue structure, \new{the span of which is larger than a single speaker-turn, but smaller than a complete
dialogue activity} \citep{traum-nakatani-99}, that is specifically designed for multi-floor dialogue. These Dialogue Structure annotations are used to make explicit the patterns of dialogue---how subsequent utterances address previous utterances or establish a new intention.  The annotations are used for both theoretical analysis as well as training data for a dialogue system, the latter of which is discussed in Section~\ref{ssec:dialogue-system}.

The scheme includes both \textit{transaction units} (TUs) and \textit{relations}. A TU is a cluster of utterances which may be from multiple interlocutors and span multiple conversational floors that together contribute to the realization (or attempted realization) of a single intent. Relations describe the relationships between individual utterances within the TU.
The scheme focuses on clustering utterances from multiple
speakers and floors into minimal units of expression and completion of intent, as well as relationships between individual utterances within the TU. 
While there are standard annotation schemas for both dialogue acts \new{(ISO 24617-2, \citep{bunt2012iso,bunt-EtAl:2020:LREC})} and discourse
relations \new{(ISO 24617-8, \citep{prasad2015semantic})}, these schemas do not fully address the issues of dialogue structure.
Of particular interest to us, and not previously addressed in other schemas, are cases in which TUs and relations span across multiple conversational floors.

Stepping back from the SCOUT corpus in particular, we note that dialogues can be characterized by distinct information states ~\citep{TraumLarsson03}. These include sets of participants, participant roles (e.g.,~active, ratified participant vs. overhearer), turn-taking or
floor-holding, expectation of how many participants will
make substantial contributions at a time \citep{edelsky_1981}, and other factors. Often distinct dialogues with different information 
states are going on at the same time.  There are a number of ways in which such
dialogues can be related to each other, 
including:
\begin{itemize}
\item having the same purpose but distinct participants (e.g., teams
  competing in a trivia contest to come up with the answer first).
\item being co-located such that participants in one can observe and
  possibly comment on the other, such as groups of people sitting at different tables at a restaurant.
\item having one or more (but not all) participants in common, where
  such participants are {\em multi-communicating}
  \citep{reinsch2008multicommunicating} (e.g.,~someone in a meeting is
  texting with one or more people outside the meeting).
\end{itemize}
In the  multi-communicating case, the multiple dialogues that a multi-communicator is part of might involve completely separate topics or be more closely related, such that satisfaction of the goals of one depends on actions in the other. For example,  a question arising in a meeting might be conveyed and answered over the
text channel. In discussing SCOUT, we use the term {\em multi-floor dialogue} to refer to cases in which the high-level dialogue purposes are the same, and some content is shared, but other aspects of the information state, such as the participant structure and turn-taking expectations, are distinct.  Situations of distributed decision-making and action are quite common (e.g., ~in a restaurant  where some people take the customer's order and others make the food; or in military units, where orders are relayed through the chain of command). In some cases, where all parties can hear all communication, we can view this as multi-party dialogue within a single floor, but in other cases not all the communications
are available to all participants---this is a case of multi-floor dialogue. We are particularly interested in capturing the latter case.\\

\subsection{Development and Refinement of Annotation Guidelines}
Like the development of the speech act extensions to Dialogue-AMR, described in Section~\ref{sec:amr}, the development of the dialogue structure annotation schema was 
iterative in that 
annotation was conducted subsequent to each SCOUT data collection experiment, which resulted in revisiting and 
updating annotation procedures to better handle novel dialogue phenomena observed in each experiment.
The original 1.0 guidelines are described at a high level in \cite{traum2018dialogue}.  
These were developed on the first 60 dialogues collected in SCOUT
in a process of single annotation and validation of 
each annotated file, where patterns of annotation errors and gaps in annotation coverage were discussed in biweekly meetings.   
Once finalized, detailed annotation guidelines were then documented in \cite{bonial2019dialogue}. 
These guidelines were applied to the annotation of the remainder of SCOUT dialogues.
Again, all annotation files
were checked in a process of validation after annotation and disagreements and challenges were discussed in 
annotation team meetings and one week-long annotation ``boot camp.''  Subsequently, 
the annotation guidelines were revised again to the current state, described next. 

\subsection{Annotation Procedure Overview}
We annotate two aspects of Dialogue Structure at the meso-level.
First, we look at {\em intentional structure} \citep{GroszSidner86}, consisting of units of
dialogue utterances that all have a role in explicating and addressing
an initiating participant's intention. Second, we look at the
relations between different utterances within this unit, which reveal
how the information state of participants in the dialogue is updated
as the unit 
progresses.
Each of these annotation levels are described in
turn below. 

\subsubsection{Transactional Units}
\label{ssec:TU}
We call the main unit of intentional structure a {\em transaction unit}, following \cite{Sinclair75} and \cite{carletta-etal:1996}.
 A transaction unit (TU) contains an initial message by one speaker and all subsequent messages by the same and other speakers across all floors to complete the intention. For example, a transaction may consist of an instruction initiated by one participant in one floor that is relayed by a multi-communicator to another floor, and then performed by yet another participant of the second floor, in addition to various sorts of feedback between pairs of participants. 

\new{The TU represents the 
lowest level of dialogue in which intentions are fulfilled across
speakers. In particularly complex negotiations or problem-solving, intentional structure can be recursive, such that the purpose of one segment partially contributes to the purpose of a higher-level segment \citep{GroszSidner86}.  Other types of dialogues have a flatter structure,  including transactions that contribute to an overall dialogue purpose, but with few, if any, levels in between. }

\new{Consider the simple} multi-floor dialogue structure from SCOUT with a single TU exemplified in Table~\ref{tab:minimal_tu}. Recall that there are: 
three participants (Commander (CMD), the dialogue manager wizard (DM), and
the robot navigator wizard (RN)), \new{and four distinct
message streams separated into two floors (for Commander and DM, called ``left'',  and for DM and RN, called ``right'') over which the DM multi-communicates and translates information from one floor to the other.} 
\new{This TU begins with the Commander's instruction to \textit{move forward three feet.} After an acknowledgement by the DM to the Commander, the DM translates the Commander's instruction to the ``right'' floor for the RN to execute. Finally after the RN's completion of the instruction, their completion acknowledgement \textit{done}} is translated by the DM from the RN to the ``left'' floor; additional details on the antecedent (``Ant'') and relation (``Rel'') annotation columns will be given in the Section~\ref{ssec:rels}.

\begin{table}[h!]
\centering
\begin{small}
\caption{Dialogue structure annotations for a single, simple TU. The ack- prefix indicates a type of acknowledgement (under \bf{Rel} column of Annotations). \label{tab:minimal_tu}}
\begin{tabular}{lp{0.5in}p{0.7in}p{0.7in}p{0.4in}llp{0.7in}}

\toprule
& \multicolumn{2}{c}{Left Floor}              
& \multicolumn{2}{c}{Right Floor}          
& \multicolumn{3}{c}{Annotations} \\

\cmidrule(r{1em}l{0.75em}){2-3}
\cmidrule(rl{0.75em}){4-5}
\cmidrule(rl{0.75em}){6-8}

\# & \textbf{CMD} & \textbf{DM$\rightarrow$CMD} & \textbf{DM$\rightarrow$RN} & \textbf{RN} & {\bf TU} & {\bf Ant}& {\bf Rel}  \\ 
\midrule
  
1& move forward three feet & & & & 1 & &\\ \hline
\rowcolor{gray} 2& & ok & & &1& 1& ack-wilco \\ \hline
3& & & move forward 3 feet & & 1 & 1 &translation-r-direct \\ \hline
\rowcolor{gray} 4& & \new{moving. . .} & & &1& 3& ack-doing \\ \hline
5& & & & done & 1& 3 &ack-done \\ \hline
\rowcolor{gray}6& & \new{done} & & & 1& 4 &translation-l \\  \hline
\end{tabular}
\end{small}	
\end{table}

\new{This dialogue exhibits a nearly unchanged form of translation of the Commander's instruction---differing only by the DM's normalization of the number \textit{three}---yet this passing of information across the conversational floors through the TU forms the critical backbone for more complex communication observed in the corpus, and in fact is modeled after the way a fully automated system would work. Dialogue management and robot motion are separate modules, here represented by separate human wizards (experimenters).}

Table~\ref{tab:extended_tu}
provides an example of a more complex TU with intervening clarifications. 
\new{The Commander's initial direction to the robot for where to face (\#1) is not precise enough for the DM to pass along to the RN. The DM describes what the robot can see---more than one doorway---and so is indirectly requesting a clarification of the Commander (\#3).\footnote{\new{ID \#3 cuts off unexpectedly: \textit{I see a doorway ahead of me on the right and a doorway.}  This is a mistake in the response sent by the DM-wizard during the early stages of collecting the SCOUT corpus, where the DM-Wizard was free-typing responses in real-time to the Commander. Later stages of data collection automated the DM-wizard responses with an interface that sped up DM responses and reduced errors \citep{bonial2017laying}}.} The Commander responds by clarifying which doorway in a repair turn to the DM (\#4). With this extra information, the DM passes along a reformulated navigation instruction for the robot to the RN (\#5).}

Each utterance-level message is assigned to at most
one TU, and the TU is defined by the set
of constituent utterances. 
At most points in a dialogue, there is only one active TU. However there are occasions where there are multiple active TUs, with a new one started before the previous one has been completed. \new{Table~\ref{tab:multiple_tu} shows an example of interleaved TUs between utterance IDs \#5 and \#9. The RN's \textit{done} in \#7 refers to the completion of \#1--2 (\textit{go into the room in front of you / and face south}) which was translated in \#3--4 (\textit{move into room in front of you / face south}). Although the Commander had already issued a new instruction in \#6 (\textit{take a picture}) it was not translated to the RN until after the RN reported completion of the movement instructions, thereby still belonging to TU 1. The image request began a new TU and was translated in \#9 ( \textit{image}), and \#10 from the RN (\textit{sent}) refers to the completion of the photo request.}

\begin{table*}
\centering
\begin{small}
\caption{Dialogue structure annotations for a single, complex TU.
\label{tab:extended_tu}}
\begin{adjustbox}{width=0.95\textwidth}
\begin{tabular}{lp{1.1in}p{1.1in}p{1.1in}p{0.2in}llp{0.8in}}
\toprule
& \multicolumn{2}{c}{Left Floor}              
& \multicolumn{2}{c}{Right Floor}          
& \multicolumn{3}{c}{Annotations} \\
\cmidrule(r{1em}l{0.75em}){2-3}
\cmidrule(rl{0.75em}){4-5}
\cmidrule(rl{0.75em}){6-8}
\# & \textbf{CMD} & \textbf{DM$\rightarrow$CMD} & \textbf{DM$\rightarrow$RN} & \textbf{RN} & {\bf TU} & {\bf Ant}& {\bf Rel}  \\ 
\midrule
1& face the doorway on your right in front of you & & & & 1 & &\\ \hline
\rowcolor{gray} 2& and take a picture & & & & 1 & 1 & continue \\ \hline
3& & I see a doorway ahead of me on the right and a doorway & & & 1 & 1 & req-clar \\ \hline
\rowcolor{gray} 4 & the one closest to you & & & & 1 & & clar-repair \\ \hline
5 & & & move to face the hallway opening to the right & & 1 & 4* & translation-r-contextual \\ \hline
\rowcolor{gray} 6 & & & image & & 1 & 2 & translation-r-direct
\\  \hline
7 & & executing... & & & 1 & 4* & ack-doing\\  \hline
\rowcolor{gray} 8 & & & & done & 1 & 6* & ack-done \\  \hline
9 & & sent & & & 1 & 8 & translation-l \\  \hline
\end{tabular}
\end{adjustbox}
\end{small}	
\end{table*}

\begin{table*}[h!]
\centering
\caption{Dialogue structure annotations for two interleaved TUs.
\label{tab:multiple_tu}}
\begin{adjustbox}{width=0.95\textwidth}
\begin{tabular}{lp{1.1in}p{1.1in}p{1.1in}p{0.2in}llp{0.8in}}
\toprule
& \multicolumn{2}{c}{Left Floor}              
& \multicolumn{2}{c}{Right Floor}          
& \multicolumn{3}{c}{Annotations} \\
\cmidrule(r{1em}l{0.75em}){2-3}
\cmidrule(rl{0.75em}){4-5}
\cmidrule(rl{0.75em}){6-8}

\# & \textbf{CMD} & \textbf{DM$\rightarrow$CMD} & \textbf{DM$\rightarrow$RN} & \textbf{RN} & {\bf TU} & {\bf Ant}& {\bf Rel}  \\ 
\midrule
1& go into the room in front of you  & & & & 1 & &\\ \hline
\rowcolor{gray} 2& and face south & & & &1& 1& continue \\ \hline
3 & & & move into room in front of you & & 1 & 1 & translation-r-direct
 \\ \hline
\rowcolor{gray} 4& & & face south & & 1& 2 & translation-r-direct
 \\ \hline
5 & & executing... & & & 1 & 2* & ack-doing \\ \hline
\rowcolor{gray} 6& take a picture & & & & 2 & & \\ \hline
7 & & & & done & 1 & 4* & ack-done\\ \hline
8 & & done & & & 1 & 7 & translation-l \\ \hline
\rowcolor{gray} 9 & & & image & & 2 & 6 & translation-r-direct \\ \hline
10 & & & & sent & 2 & 9 & ack-done\\ \hline
11 & & image sent & & & 2 & 10 & translation-l \\ \hline 
\end{tabular}
\end{adjustbox}
\end{table*}

\subsubsection{Relations \& Antecedents}
\label{ssec:rels}
We model the internal structure of TUs as {\it relations}
between pairs of utterances within the unit. Each relation is annotated by coding a relation-type and an antecedent for each utterance starting after
the first utterance in a transaction. Thus, each transaction unit can be viewed as a tree structure, with the first utterance as root (having
no relation-type or antecedent annotations). While relations often exist between
an utterance and multiple previous utterances, to simplify the annotation, we code only the
most direct, recent such relation. This practice is common for
many annotation efforts (e.g., the ``code-high'' principle from~\cite{CondonCechMan}).
In the future, we plan to use
inference rules to derive  ``indirect relations''  from what has
been annotated.

Relations are organized into a taxonomy of types, \new{representing how each utterance relates to its antecedent. This covers some of the same phenomena addressed by taxonomies of discourse relations (e.g., \cite{prasad2015semantic}), as well as backward looking dialogue acts \cite{Damsl} and several dimensions of the ISO standard for dialogue act
annotation \cite{bunt2012iso,bunt-EtAl:2020:LREC}. However it also covers relations across multiple floors and status updates, which are not well represented in prior schemes.
The highest-level distinction of the taxonomy relates to speaker and floor structure.
\textsc{expansions} are relations between utterances of the same speaker and within the same conversational floor. \textsc{responses} are relations between utterances by different speakers within the same floor. \textsc{translations} are relations between utterances in different conversational floors. Within each of these broad relation types, there is at least one but often two levels of relation subtypes.}

\new{Expansions correspond to the same subject matter as discourse relations, however we only consider a few relation types to distinguish new, replaced, and redundant information and turn management rather than the fine internal structure of arguments. Most of the relations from the ISO discourse relations scheme are coded here as ``continue". Future work may involve adding additional relation subtypes to capture argumentation structure.}  

\textsc{translations} have two main subtypes to characterize \new{how} the information is being translated.
\new{Translation of information from the left floor to the right floor occurs when the DM passes the Commander's instructions to the RN. It appears in the DM$\rightarrow$RN stream and is called a \textsc{translation-\underline{r}ight}. Information from the right floor to the left floor occurs when the DM passes the RN's responses to the Commander. It appears in the DM$\rightarrow$CMD stream and is called a} \textsc{translation-\underline{l}eft} (see Tables~\ref{tab:minimal_tu},~\ref{tab:extended_tu}, and~\ref{tab:multiple_tu}). \new{There are also categories where only a part of the intention is translated or when the speaker relays or talks about issues in the other floor rather than translating the same intention.}

\new{In order to support features of situated dialogue, the \textsc{translation-right} relations were assigned subtypes to indicate if the response was appropriate only given a particular situated, physical setting \citep{bonial2021context}. An example of this is in \textit{Go to the door ahead}, where the instruction being translated directly references the physical environment the robot saw at the moment the instruction was issued.}
The door specified in the translation of such instructions will change depending upon the robot's current physical environment, 
\new{and in this case would be a \textsc{translation-right-situated} relation. Specifically labeling these instructions paves the way for }
mapping to robot plans that are valid in the current environment.

\new{\textsc{response} corresponds roughly to the backward looking acts of \cite{Damsl} and several dimensions of the ISO scheme. All of these relations provide some sort of feedback~\citep{Allwood92}, as to how the responder has perceived, understood, and reacted to the antecedent. The feedback can be positive or negative at multiple levels}. While some of these relations correspond to higher-level backward-looking acts, like accept and offer, or answer a question, others refer to the grounding process~\citep{Traum94d} and conveying the planning and execution status of an instruction.

\new{The finer granularity of \textsc{acknowledgment} response types enables modeling relatively nuanced feedback as to the robot's confidence in the understanding of the instruction. For example, } \textsc{ack-doing} and \textsc{ack-done}, indicate that an instruction is being or has been carried out \new{while \textsc{will-comply} expresses full confidence that the robot has understood and will execute the instruction.  The \textsc{unsure} and \textsc{try} subtypes, in contrast, portray an acknowledgment of what was understood and a lack of confidence, inviting repair from the Commander if needed.} See Table~\ref{tab:annotation_relations} for a 
full list of relations. 

\begin{table}[h!]
  \centering
    \caption{Annotation Relations}
  \begin{tabular}{llp{1.5in}}
    \toprule
    \textbf{General Relation Type} & \textbf{Relation} & \textbf{Relation Subtype} \\
    \midrule
    \multirow{4}{*}{Expansion} &  Continue &  \\ 
    & Correction &  \\
    & Link-next &  \\ 
    & Summarization & \\ \midrule
    \multirow{5}{*}{Translation} & Translation-left & \\ \cmidrule{2-3}
    & Translation-right & direct, contextual, landmark, situated, history, default \\ \cmidrule{2-3}
    & Partial translation & Translation-X partial \\ \cmidrule{2-3}
    & Quotation &  \\
    & Comment &  \\
    \midrule
    \multirow{21}{*}{Response} & Processing &  \\ \cmidrule{2-3}
    & Acknowledge & underspecified, understand, unsure, try, will comply, will do prep, doing, doing prep, done, can't \\ \cmidrule{2-3}
    & Partial acknowledgment & partial \\ \cmidrule{2-3}
    & Clarification & request, repair, repeat, done status\\ \cmidrule{2-3}
    & Request & repeat, done status \\ \cmidrule{2-3}
    & Question-Response & answer, non-answer response \\ \cmidrule{2-3}
    & Offer accept & \\ 
    & Offer reject & \\ 
    & Reciprocal response & \\ 
    & Third-turn feedback & \\ 
    & Other response & \\
    \bottomrule
  \end{tabular}
  \label{tab:annotation_relations}
\end{table}

While relation annotations specify the relations between one utterance and another previous utterance, the antecedent annotation marks up precisely which past utterance an annotation target is related to in the manner specified by the 
relation. For example, in Table~\ref{tab:extended_tu}, \#3 \new{(\textit{I see a doorway ahead of me on the right and a doorway})} is a request for a clarification of \#1 \new{(\textit{face the doorway on your right in front of you})}, as specified in the ``Ant'' (antecedent) and ``Rel'' (relation) columns. 

For multiple commands in succession by the same speaker and part of the same group, each line has the preceding line as its antecedent. For an utterance that is directly related to a whole sequence of utterances from the same speaker, we use the last line of that sequence along with an asterisk.
Again in Table~\ref{tab:extended_tu}, the translation in \#5 \new{(\textit{move to face the hallway opening to the right})} is of the entire complex instruction and clarification sequence from \#1--\#4 \new{(\textit{face the doorway on your right in front of you / and take a picture / I see a doorway ahead of me on the right and a doorway / the one closest to you})}; this is specified through the use of the 4* antecedent label, and the connection between \#1 \new{(\textit{face the doorway on your right in front of you})} and \#2 \new{(\textit{and take a picture})} is specified through the \textsc{continue} relation between them and the subsequent \textsc{clarifications}.

\pagebreak

\subsection{Dialogue Structure Corpus Summary}

All 89,056 utterances in SCOUT have been annotated for dialogue structure. 
As the annotation guidelines were updated in several iterations, we 
measured IAA at two major points in the annotation development: 
after the development of the first set of guidelines described in 
\cite{traum2018dialogue} (i.e., the unmodified 2018 schema), and after the most major revamping of 
the guidelines to better handle situated dialogue described in \cite{bonial2021context} (i.e., the modified 2021 schema). 
We compute IAA on the three markables in the annotation schema: antecedents, relations, and transaction units (TUs). Three expert coders annotated a subset of 3 dialogues (a total of 896 utterances) using the modified (2021) schema. Results appear in Table~\ref{table:inter-annotator-agreement}, which also shows  the reported IAA from the unmodified (2018) schema. Note that in the unmodified schema, two rounds of IAA were conducted, the first round on 3 dialogues of 482 utterances using 5 coders,
and the second round on a single dialogue of 314 utterances using 6 coders. We compare this range of IAA from the four trials of the unmodified schema, to the range of IAA for the three trials annotated with the new schema. 

\begin{table}[h]
\small
  \caption{IAA of the original, unmodified schema of \cite{traum2018dialogue} and our modified schema.}
  \begin{tabular*}{\columnwidth}{@{\extracolsep{\fill}}lccc}
    \toprule
    \raisebox{-1.87ex}[0pt][0pt]{Markable Type} 
    & \multicolumn{2}{c}{Agreement (Krippendorff's $\alpha$)}
    & \raisebox{-1.87ex}[0pt][0pt]{\begin{tabular}{@{}c@{}}Distance\\Metric\end{tabular}} \\
    \cmidrule{2-3}
    & Unmodified Schema          & Modified Schema     &   \\
    \midrule
    Antecedents       & 0.72--0.82 & 0.79--{\bf 0.94} & Nominal \citep{krippendorff:80} \\
    Relation Types    & 0.77--0.89 & 0.83--{\bf 0.93} & Nominal \citep{krippendorff:80} \\
    Transaction Units & 0.48--{\bf 0.93} & 0.65--0.85 & MASI \citep{passonneau:2006:LREC} \\
    \bottomrule
    \multicolumn{4}{@{}l@{}}{\small
    }
  \end{tabular*}
  \label{table:inter-annotator-agreement}
\end{table}%
Our modified schema yields comparable or higher IAA than the original schema for antecedents (maximum 0.94) and relation types (maximum 0.93). Our TU IAA (maximum 0.85) is higher than the range of TU IAA reported for the first round of annotations with the unmodified schema (0.48--0.70), but the final round of TU annotation from in the unmodified schema achieves the highest agreement rate of 0.93. Note that our modified schema adapts the same coding for antecedents and TUs. Thus, although one might expect that adding annotation categories would lead to lower IAA, the addition of our new subtype relations did not produce significantly lower agreement scores, demonstrating that the new annotation categories are clearly identifiable.

\subsection{Dialogue Structure Annotation Summary}

To support dialogue system development in complex, situated tasks, there 
needs to be a strong cohesion between the dialogue and the surrounding 
physical environment. Specifically, instructions sometimes can 
only be interpreted sensibly when interpreted with respect to the physical 
surroundings. For example, an utterance that is, on its face, a question---``Can you move forward 2 feet?''---when interpreted with respect to the 
current physical environment may be a question of ability (i.e., is there 
room to move forward that far?) or may be a politely worded command. 
\new{Excerpt \hyperref[excerpt2]{2}} is shown again in Table~\ref{tab:excerpt1-ds}, now with \new{details on each conversational floor 
and} the Dialogue Structure 
\begin{table*}[t]
\centering
\begin{small}
\caption{\new{Excerpt \hyperref[excerpt2]{2}} shown with conversational floors and Dialogue Structure TU, Antecedent 
and Relation annotations}
\label{tab:excerpt1-ds}
\begin{adjustbox}{width=0.95\textwidth}
\begin{tabular}{lp{0.9in}p{0.85in}p{0.5in}p{0.4in}llp{0.7in}}

\toprule
& \multicolumn{2}{c}{Left Floor}              
& \multicolumn{2}{c}{Right Floor}          
& \multicolumn{3}{c}{Annotations} \\

\cmidrule(r{1em}l{0.75em}){2-3}
\cmidrule(rl{0.75em}){4-5}
\cmidrule(rl{0.75em}){6-8}

\# & \textbf{CMD} & \textbf{DM$\rightarrow$CMD} & \textbf{DM$\rightarrow$RN} & \textbf{RN} & {\bf TU} & {\bf Ant}& {\bf Rel}  \\ 
\midrule
  
1& robot continue down the hallway directly in front of you underneath the overhead light& & & & 1 & &\\ \hline
\rowcolor{gray} 2& & I don't see an overhead light in my current position.  Would you like me to send a photo? & & &1& 1& offer \\ \hline
3& robot send a photo & & & & 1 & 2 &offer-accept \\ \hline
\rowcolor{gray} 4& & & photo & & 1& 3 &translation-r-direct \\ \hline
5& & & & image sent& 1& 4 &ack-done \\  \hline
\rowcolor{gray} 6& robot continue moving forward to the right of the red bucket& & & &2 & &\\ \hline
\end{tabular}
\end{adjustbox}
\end{small}	
\end{table*}
annotations. Note that the DM is unable to translate
the initial instruction due to the misunderstanding, to which the DM \textsc{offers} to send a photo, and the Commander accepts. The newly formed instruction is translated within the same TU, and a new intent to move to the red bucket begins a new TU. \new{Refer to Table~\ref{tab:amr-ds} to see the Dialogue Structure annotations alongside the Dialogue-AMR annotations for this excerpt; the two annotation schemas are complementary in providing both within utterance meaning and intention, as well as the relation that each utterance bears to another.} See Section~\ref{ssec:dialogue-system} for a discussion of how these annotations have been used to implement a dialogue management system.

\newpage

\section{Visual Context Annotations}
\label{sec:multi-modal}

Here, we describe ongoing annotation efforts to characterize the relationships in SCOUT between the language and the visual affordances: images and LIDAR. The DM had access to the same visual information as the Commander in addition to a real-time view into the environment, all of which were necessary for the DM to determine the intent of the Commander's utterance and how to respond based on the current environment and the robot's position within that space.

\subsection{Annotation Description}

We describe two annotation tasks that form building blocks for understanding the visual context and reasoning required for the Commander and the DM to complete their collaborative task. First, we treat and analyze the sent images as static, shared snapshots into the robot's environment. Second, we treat the LIDAR map as an evolving and persistent shared common ground. As these efforts are the most recent in our body of work, we describe the motivation and results of our annotations, followed by planned work to expand annotation.

\subsubsection{Photo Requesting Strategies}

The images provide the Commander with a semi-persistent snapshot into the environment---the environment itself does not move, but as soon as the robot changes position, it will no longer be oriented in the same direction as when it took its most recent snapshot. The images also provide the DM with the Commander's knowledge about the space. The Commander only knows what they see through the images, which can help the DM to establish and correct common ground. 
We begin our analysis by identifying how these snapshots were obtained in annotating dialogues for photo requesting strategies. We focused on requests that were {\it Commander-initiated} in order to establish a baseline for normative Commander behavior in the absence of miscommunication.

The Commander's utterances were analyzed by one annotator who read through the dialogue and, for each image taken, traced backward in the dialogue to discover the moment the request was initiated and by whom. The following categories were identified as Commander-initiated requests:

{\bf Front} strategies requested that a single, forward-facing photo be taken, typically after the robot completed an activity, for instance:
\begin{itemize}
    \item \textit{``move forward five feet then take a photo''}
    \item \textit{``robot proceed through the doorway right in front of you and take a photo''}
    \item \textit{``go to the wall behind you. face north. and then take a picture''}
\end{itemize}

{\bf Cardinal} strategies requested that four photos in the cardinal directions be taken, for instance:
\begin{itemize}
\item \textit{``take pictures in north south east and west directions''}
\item \textit{``move forward to the middle of the room and take pictures in north east south and west direction''}
\end{itemize}

{\bf Degrees of rotation} strategies requested that a sequence of photos in a circle, semi-circle, or quarter-circle be taken, for instance:
\begin{itemize}
\item \textit{``pivot three hundred and sixty degrees to the right taking a picture every forty five degrees''}
\item \textit{``pivot one hundred and eighty degrees to the right taking a picture every forty five degrees''}
\item \textit{``rotate ninety pivot ninety degrees right taking a picture every forty five degrees''}
\end{itemize}

A {\bf Repetition} strategy was observed in which the Commander asked the robot to fulfill their desired picture taking strategy after their every subsequent command without them having to explicitly state it. While each subsequent fulfillment of the request was performed by the robot without explicit instruction from the Commander, the original request was still Commander-initiated, for instance: 
\begin{itemize}
\item \textit{``move forward into the room and take pictures after each movement''}
\item \textit{``take pictures in all four directions after each movement''}
\end{itemize}

These agreements lasted until the trial ended or could be terminated by the Commander at any time,  e.g., \textit{``do not take pictures after new movements''}.

\subsubsection{LIDAR Exploration Maps}

While the images provide snapshots into the robot's current environment, the LIDAR maps provide a real-time view of large obstacles and the robot's position and orientation in its explored floor plan. The location of the robot is overlaid on the LIDAR map with its orientation as well, so the Commander can see at all times where the robot is and which way it is facing. However, in the same way that the images are only available upon request, the LIDAR map too only expands as the Commander directs the robot through the environment. This is a persistent and shared resource between the Commander and DM, so the DM knows at every moment what the Commander sees on their LIDAR map, and takes this into consideration when determining the success and execution of instructions issued. 

To build up to this real-time robot interpretation of the LIDAR maps for interpreting instructions, we begin by constructing a single Environment Map per dialogue, rather than an evolving instruction-by-instruction map. By first understanding the environment at a single moment in time, we pave the way for interpreting the environment at each time-step. 

\begin{figure*}[ht]
    \centering
    \begin{subfigure}[t]{0.34\textwidth}
        \includegraphics[width=\textwidth]{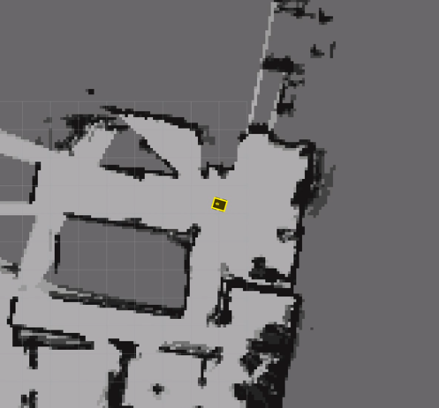}
        \caption{\label{fig:endmap}LIDAR map at the end of a dialogue. Dark gray are areas unscanned by the LIDAR. Robot icon in center of map.}
    \end{subfigure}
    \begin{subfigure}[t]{0.24\textwidth}
    \centering
        \includegraphics[width=0.8\textwidth]{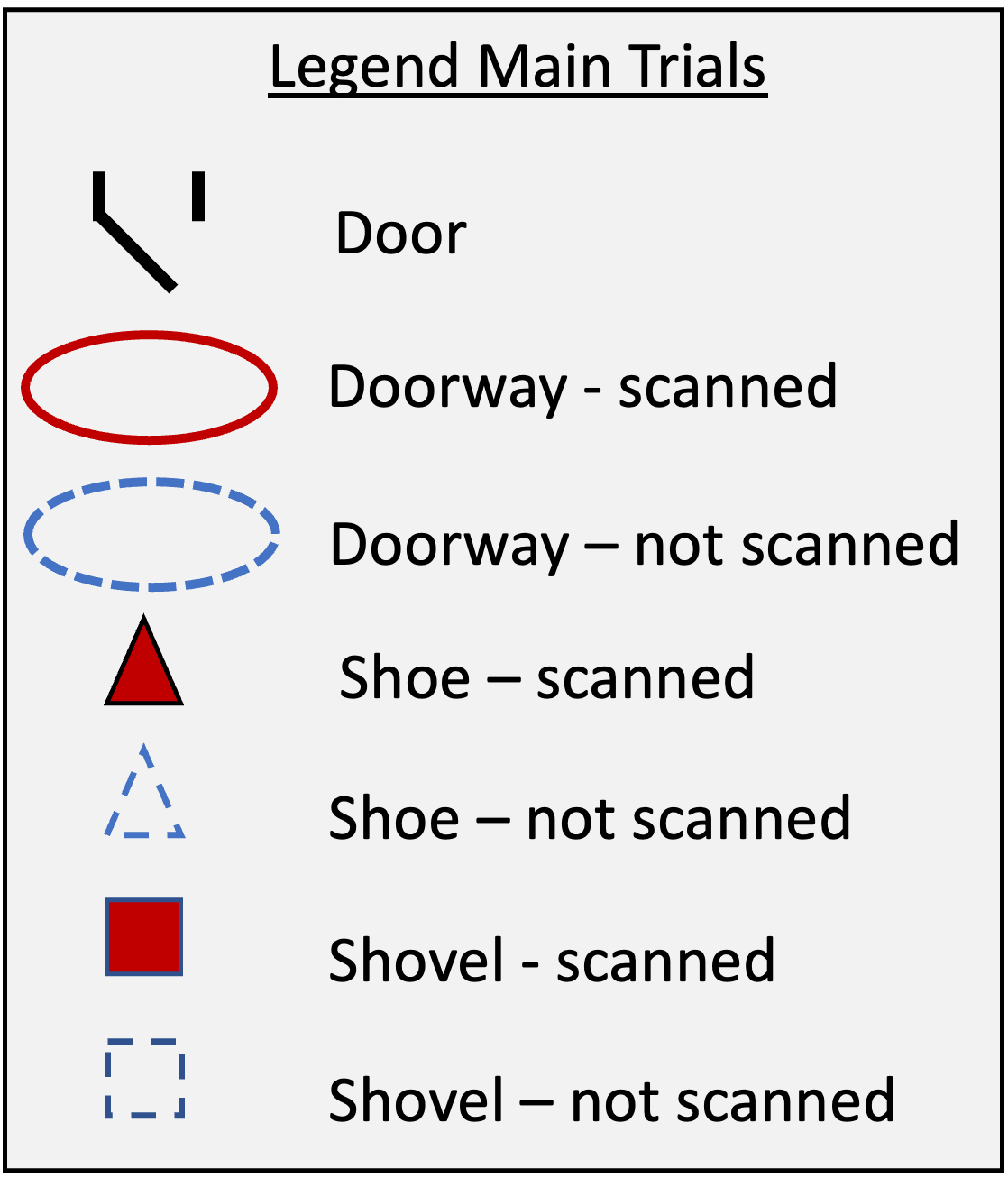}
        \caption{\label{fig:legend}Legend}
    \end{subfigure}
    \begin{subfigure}[t]{0.37\textwidth}
        \includegraphics[width=\textwidth]{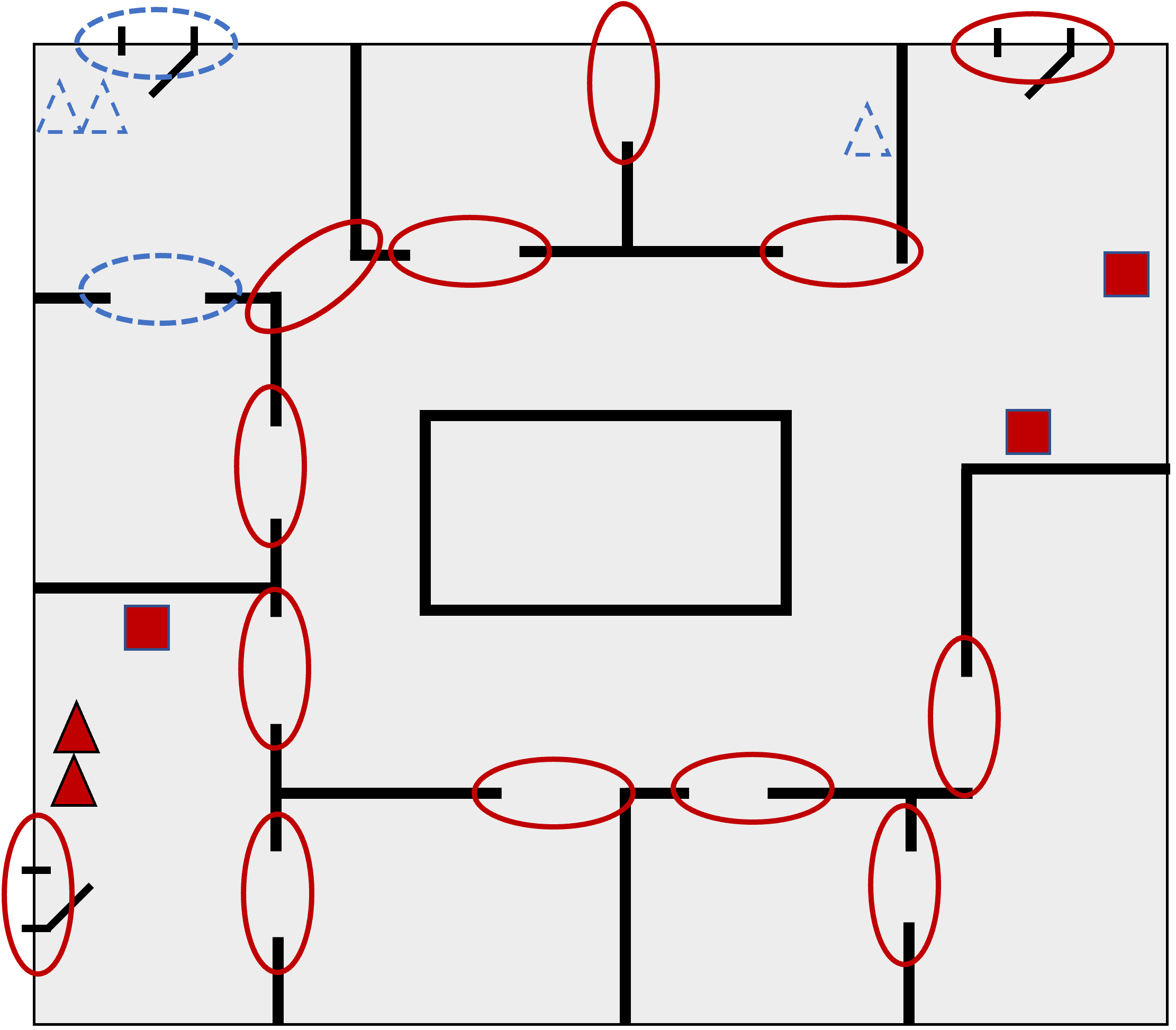}
        \caption{\label{fig:floorplan}Corresponding top-down floor plan of Fig~\ref{fig:endmap} annotated with which items were scanned by the LIDAR.}
    \end{subfigure}
    \caption{LIDAR map, legend, and  Exploration Map with items scanned or not scanned by the LIDAR}
    \label{end-map}
\end{figure*}

A LIDAR map was extracted from the last frame of the Commander's screen as captured by the screen recording of their workstation (e.g.,  Figure~\ref{fig:endmap}). From this, an annotated Exploration Map was drawn and marked up to denote all items in their known location in the environment that the Commander was tasked to search for: doorways, shovels, and shoes. The Exploration Map differentiates between items which were scanned by the LIDAR over the course of the dialogue (marked on the exploration map with solid red lines), and those that were not (marked in dashed blue (see Fig.~\ref{fig:legend} for the legend). The resulting Exploration Map corresponding with the LIDAR in Figure~\ref{fig:endmap} is shown in Figure~\ref{fig:floorplan}. 

Annotators were given the LIDAR map and an Exploration Map template with all the markers for doorways, shoes, and shovels already in place and indicated as `scanned' by default. The annotator then changed the markers on the map to the unscanned scheme if the target on the LIDAR was not scanned. This determination was based on if the LIDAR shows light gray or dark gray, where dark gray denoted areas not scanned. Annotations were conducted by one annotator, and verified by a second. In cases of uncertainty, the screen recording of the Commander's full trial was reviewed to obtain a more comprehensive overview of the path that the robot took. 

A text list of the items in the Explorations Maps was created to allow for computational processing with a unique identifier assigned to each target starting in the upper left-hand corner and moving clock-wise, and denoted as scanned or not scanned:\\

\begin{small}
\begin{verbatim}
    door1    not-scanned
    door2    scanned
    ...
    shoe1    not-scanned
    shoe2    not-scanned
    ...
    shov1    scanned
    shov2    scanned
    ...
\end{verbatim}
\end{small}

\subsection{Visual Context Corpus Summary}

Photo-requesting strategies and LIDAR Exploration Map annotation has been completed on 30 dialogues in SCOUT with plans to automate the annotation and verification process until all 287 dialogues are completed.

\subsection{Visual Context Annotation Summary}

\new{Recall from Section~\ref{sec:space} that the Commander is tasked with instructing the robot through a series of search and navigation tasks. These tasks include finding and counting doorways (which can be found more readily from the LIDAR map), cones, shovels, or shoes (which can all only be found through an image). We find that the} Commander's photo-requesting strategies \new{can be used} to analyze success in these search tasks.
Commanders who relied on the `Degrees of rotation' strategy achieved very high scores on these tasks \citep{lukin2023navigating}. Furthermore, two Commanders who  began their first main trial with a `Front' strategy later experimented with a `Cardinal' strategy, and used it in their entire second main trial, suggesting they deemed the latter to be more effective. With this strategy, these Commanders were able to obtain more snapshots, presumably becoming more informed about the environment which may have lead to higher success, or at the very least, higher satisfaction with the chosen photo strategy. The Exploration Maps have been used to compute measures of success showing that increased exploration yielded higher task success scores \citep{lukin2023navigating}.

Returning to the dialogue and images from \new{Excerpt \hyperref[excerpt2]{2}}, we observe that the miscommunication occurs due to the Commander's misunderstanding about where the robot is and what it sees in the environment. The \textit{overhead light} that they saw in the last photo requested was not consistent with the robot's current position. Here, it was the DM who noticed the discrepancy between the robot's view and intention of the instruction, and so initiated a repair strategy. Only the DM could do so by being able to understand the Commander's intent and how it (the robot) should respond. In this case, the Commander's photo-requesting strategies may have been insufficient in continually providing them with common ground, but the strategies did provide the DM enough information to recognize the discrepancy and mitigate it. 

See Section~\ref{ssec:visual} for a discussion of how we envision using these \new{visual context annotations of images and maps} in dialogue management, and further challenges involved in sharing visual contextual information.

\section{Discussion: Annotation and Systems}
\label{sec:use-case}

\begin{figure*}[ht]
\centering
\includegraphics[width=4.6in]{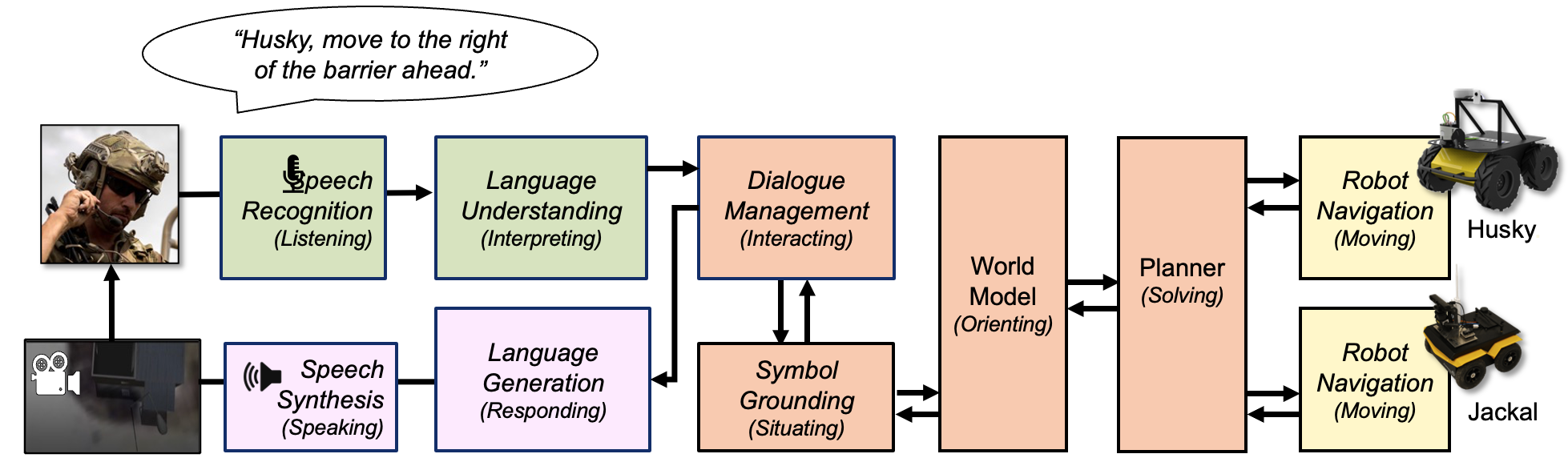}
\caption{System architecture, supporting bi-directional, grounded communication between a Commander and multiple remotely located robots.}
\label{fig:architecture}
\end{figure*}

In developing our annotation schemas and labeling the SCOUT dataset, we are moving towards understanding how humans would naturally speak to robots in the remote, collaborative paradigm, and how robots should respond in turn. We outline one aspirational framework for a dialogue system to run onboard a robot in Figure~\ref{fig:architecture}.  The framework touches upon necessary components and capabilities, including {\it language understanding} of the human speech and interpreting their intent; {\it dialogue management} to 
decide how to respond appropriately, whether it be moving in a physical space, providing a status update, or clarifying an unclear instruction; and {\it grounding} to map the language terms to entities and information about the physical world identified by robot’s sensors in physical world, requiring a robot to reason over multiple sources of uncertainty: language, speech, vision, occupancy, knowledge. A framework such as this would enable robust dialogue with robots. In the remainder of this section, we discuss our progress towards implementing such a framework harnessing our annotations. We show evidence of successful integrations, and identify ongoing needs and use-cases of the annotations.

\subsection{Dialogue Management for Autonomous Robot Navigation Systems}
\label{ssec:dialogue-system}
The dialogue structure TU and relation annotations were leveraged as training data 
for developing three iterations of dialogue management systems detailed in the paragraphs
to follow. All systems leveraged the dialogue structure data organized into 
instruction-response pairs, where instructions were previously-seen Commander instructions, and responses were either messages sent back to the Commander in the DM$\rightarrow$CMD stream, such as 
feedback or clarification questions, or messages sent on to the robot in the DM$\rightarrow$RN stream
for further processing and execution.  
The training data instruction-response pairs were used to learn the weights of association such that a ranked list of potential matches was returned and the most similar instruction-response pair was selected \citep{leuski2011npceditor}.  
The resulting dialogue managers were adapted from the Virtual Human 
Toolkit \citep{hartholt-EtAl:2013:IVA}, refined to support a robot platform \citep{marge2016applying}. 
The dialogue management systems have two basic elements: first, a classifier that interprets the language with respect to the basic intent, and second, a dialogue policy manager that dictates 
what the system should do next.

Intent classification, for the first element, is treated in this work as a retrieval problem, such that given the transcribed 
speech from the recognizer, the system can infer the intent by retrieving the 
\new{closest known intent, given a relevance model translation learned from the dialogue structure instruction training data. In this way the classifier can react to previously unseen but similar expressions that convey the same intent.}
For example, if the Commander provides the 
instruction \textit{Okay, Husky, check the path in front of you}, 
the system retrieves  \textit{Scout the path in front}.

Dialogue management policies, for the second element, are defined based upon the matches obtained from the intent
classifier, with two basic categories of response policies.  The first is for actionable 
messages, where the robot is able to execute the instruction.  For actionable commands, 
the basic policy is two-fold, both to respond to the Commander with feedback, demonstrating 
successful receipt of the instruction, and to send a simple text message of the 
instruction on to the robot software stack. \new{In the above example, the system would provide feedback like \textit{executing} and} pass the text instruction  along to the parsing component operating within the software stack for processing and eventual execution.
The second policy is for non-actionable messages, which requires clarification through further dialogue. The basic policy for non-actionable messages is to prompt the Commander for clarification, such that any inability to infer the intent of the instruction 
can be resolved promptly through dialogue.  
Sub-policies for non-actionable 
messages leverage the dialogue structure features 
in the annotated corpus training data to define a dialogue management policy.
For example, there is a sub-policy defining how to 
respond to incomplete instructions lacking a clear end state.
To illustrate this policy, a command from the Commander such as \textit{Keep moving forward}, would generate a query for clarification (e.g., Robot:\textit{Where should I move forward to?}; yielding a response from the Commander: \textit{To the door on the left}).
What is then passed along for execution is a complete instruction (e.g., \textit{Move forward to the door on the left}).

Our first dialogue system prototype, {\bf ScoutBot}, was created to
determine if the data collected in SCOUT could be
used to train a dialogue system to support collaborative navigation in
a similar domain~\citep{Lukin-EtAl:2018:ACL}. ScoutBot
was the first human-robot dialogue system trained entirely from data collected
using the data-driven Wizard-of-Oz methodology. ScoutBot implemented a shallow pass through the Figure~\ref{fig:architecture} framework, and permitted users 
to issue verbal navigation instructions to a virtual Clearpath Robotics
Jackal in an indoor environment with a custom ROS wrapper for robot navigation of metric instructions (e.g., \textit{Drive forward 10 feet; rotate left 45 degrees; take a photo}). ScoutBot used Google ASR to support speech recognition. Of note, ScoutBot lacked grounding, a world model, and speech synthesis; language generation was retrieval based.

Our second prototype, {\bf MultiBot}, extended ScoutBot capabilities
to support dialogue interaction with multiple robotic platforms, and in
a different task domain~\citep{marge-etal-2019-research}. By combining
dialogue with robotic behaviors (\textit{Tactical Behavior
Specifications}~\citep{boularias2015grounding,holder2017}), 
MultiBot could interpret
goal-based instructions (e.g., \textit{Robot, scout Route Bravo ahead}) to a heterogeneous, aerial-ground
team of robots based on each robot's capabilities in a search task.
As a successor to ScoutBot, MultiBot
demonstrated the generalizability of the
technical contributions to now enabling dialogue processing between
one human and a team of mobile robots. MultiBot additionally integrated speech synthesis, yet also still did not support grounding or world modeling. A new dataset of instruction-response pairs was curated to support the new domain. 

Finally, the dialogue management prototype developed in ScoutBot and MultiBot was later integrated into our robot front-end application 
called {\bf JUDI}, the Joint Understanding and Dialogue Interface \citep{marge2023bot}, 
which features integration with offline speech recognition provided by the Kaldi open-source speech recognition toolkit~\citep{povey2011kaldi}. In contrast to many of today's conversational systems, JUDI does not require a cloud connection for ASR functionality, making it suitable for use in search and rescue and disaster-relief operations where internet connectivity is unreliable or unavailable. JUDI is agnostic to robot specification, and can be adapted to different suites of navigation and perception algorithms compatible with multiple robot platforms. 

As noted, these systems lack the grounding and world model components outlined in Figure~\ref{fig:architecture}, and yet they are able to support simple human instructions for metric movement given a wide vocabulary of instruction-giving from the human collaborator. The treatment of language understanding as a retrieval problem is applicable for many of the instructions, but lacks the ability to match the order of instructions. For example, \textit{Go to the barrel then take a picture} might match on \textit{Take a picture then go to the barrel.} We seek to address these limitations with the system described in the next section. 

\subsection{Language Understanding and Grounding for Autonomous Robot Navigation Systems}
\label{ssec:amr}

In order to allow for more flexible language understanding and interpretation beyond  retrieval methods, we turn to Standard-AMR and Dialogue-AMR to ground the \textbf{meaning} of the instructions, rather than just the \textbf{words} of the instructions. This facilitates grounding and a world model (refer back to Figure~\ref{fig:architecture}).

Our grounding research is currently ongoing and being conducted in stages. We seek to compare the performance of the same grounding approach with different linguistic parses---first a simple, constituent CYK parser \citep{younger1967recognition}, then Standard-AMR, and finally Dialogue-AMR. To date, we have implemented a system for natural language control of a Clearpath Robotic Husky \citep{husky2023} using JUDI (introduced in Section~\ref{ssec:dialogue-system}) for speech recognition and dialogue management.  The architecture leverages a custom ROS wrapper around a Standard-AMR parser that parses the output of the JUDI intent classification and then passes the parse along to the planning and execution components of the autonomy stack for robot navigation. 
Our experimentation thus far demonstrates that AMR-based grounding of natural language instructions allows our system to successfully ground and execute instructions with a range of linguistic phenomena, including light verb constructions, coreference, and spatial relations. Although these phenomena are arguably complex for grounding and have proven to be challenging for existing state-of-the-art systems, they are commonplace in natural language. 
See \cite{bonial-dmr-2023} for details of the grounding research, \new{including the strategy for evaluation, which is ongoing.}

We are continuing to update our architecture so that the language understanding and dialogue management components work more synergistically with the grounding and planning components. Ideally, the language would be interpreted with an awareness of the environment in mind, prior to attempting to plan a path for executing the instruction and running into a failure. This requires cross-communication between the dialogue system and the robot's sensors, e.g., cameras or LIDAR, that can collect and draw upon knowledge of the surrounding environment to support more human-like conversational repairs in cases of ambiguities and miscommunications. For example, 
if the system encounters the well-formed instruction, \textit{Move to the barrel on the right}, it should be able to assess from its sensors the situations when there is no barrel on the right. If instead, it can determine there is a barrel on the robot's \textit{left}, that information from the grounding component can support generation via AMR, of a targeted clarification question, such as \textit{I don't see a barrel on the right; do you mean the one on the left?} This requires a level of intercommunication of the components that we currently have not achieved, and, as far as we are aware, has not been developed elsewhere.

\subsection{Shared Modalities and Participant-Specific Knowledge}
\label{ssec:visual}

\begin{figure*}[h]
\centering
\includegraphics[width=4.6in]{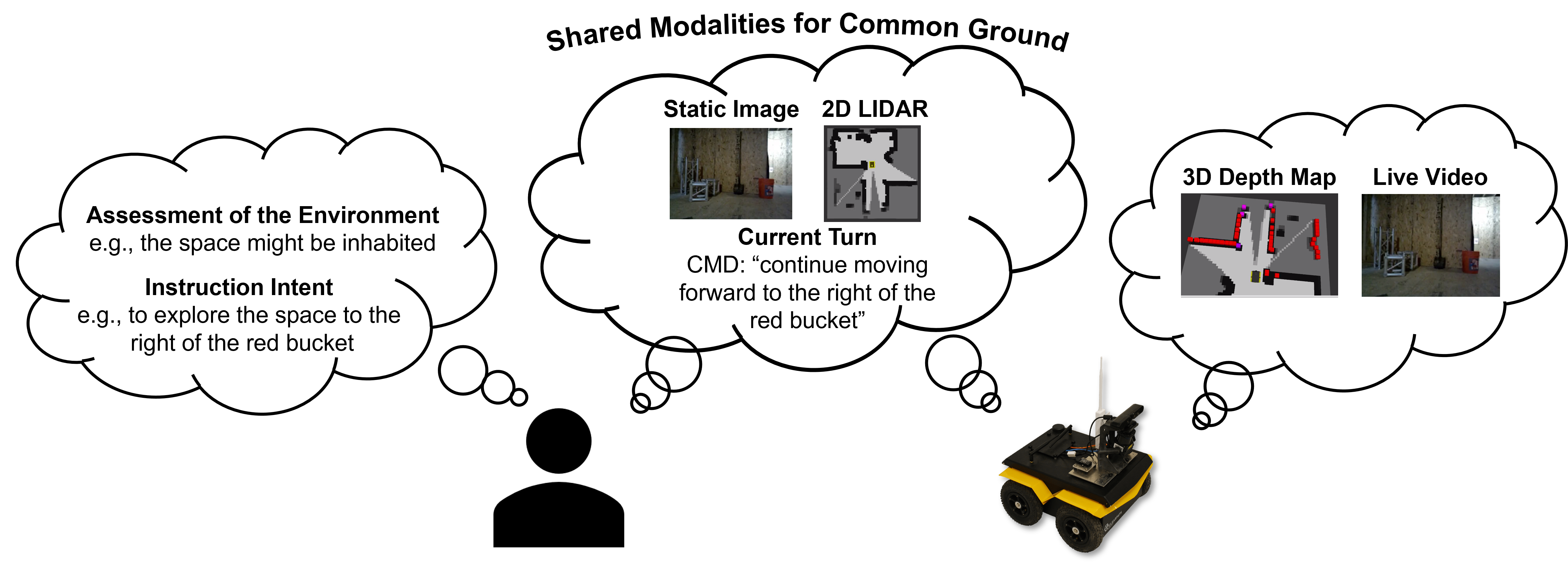}
\caption{Shared modalities for establishing common ground, along with additional information available or known to each participant during  collaboration}
\label{fig:commonground}
\end{figure*}

We envision a computational framework where shared affordances---the static photo, dynamic LIDAR map, and current dialogue turn (center of Figure~\ref{fig:commonground})---can be processed together by the robot using multi-modal annotations to help it establish common ground in real time with its teammate. While this work is ongoing, we are moving towards bringing the visual information seen by the human and recorded by the robot into the processes in Figure~\ref{fig:architecture}. The exact location in the architecture for this is beyond the scope of this discussion; instead, we present vignettes with challenges in which the shared modalities may provide signals for establishing and re-establishing common ground.

While the modalities themselves are shared, the human might not be making the same assessments of the environment as the robot. Therefore it is not enough to share the information with both the human and the robot and assume perfect knowledge and understanding between them. As in the case of \new{Excerpt \hyperref[excerpt2]{2}}, the Commander didn't recognize at first the discrepancy between the image shown to the them and the current LIDAR scan. In other cases, the human may overlook something in the image due to dark lighting in the environment, or may not recognize or recall if they had previously been to a location on the LIDAR map. 

To this end, we have been analyzing the DM-initiated photo requests and the circumstances in the dialogue which prompted the DM to ask for additional visual information. These circumstances with repair utterances, \new{as revealed from the dialogue structure annotations}, will be compared against the non-repair instances from the Commander-initiated requests as controls, to identify the point at which the common ground is lost. We hypothesize there are several instances which may lead to this, that can be learned from the annotations. Extracting and annotating moment-by-moment LIDAR maps from our collection of dialogue ROS bag files may reveal other signals for the robot to use to determine when the Commander may be in danger of losing common ground, such as when they have not requested a photo after the robot moved a significant enough distance away from the location of the prior persistent image. Identifying when this visual common ground is lost may enable the robot to \new{leverage the dialogue history from its ongoing recording of the dialogue structure to} anticipate potential repair circumstances and offer strategies to avoid them, or it may lead the robot to detect that the content of the current image does not adequately match the Commander's instruction, and preemptively provide a new snapshot.

Another challenge of establishing common ground is that fact that the human and robot each have their own additional knowledge or assumptions that may make coordination difficult. In the SCOUT experiments, the robot, for instance, was able to see live video, but it could not broadcast this signal due to the experimental design constraints on network bandwidth. Its LIDAR sensor was more complex than our currently constructed 2D Exploration Map conveys; it could sense depth of objects and obstacles around its environment (refer to  Figure~\ref{fig:commonground}). When such information is not available for the human, the robot may make assumptions based on its stored knowledge that conflict with the human's knowledge. For instance, the 3D LIDAR may detect an object low to the ground that would be not rendered on the 2D map or that would be just out of view of the robot's camera perspective. Awareness of this discrepancy of information, however, can lead to initiations of repair on the robot's part. Similarly, the human, while conducting their task, may formulate an evolving view of the environment as they observe and explore. This may not be conveyed out loud to the robot, and so the robot may attempt to resolve a misstep of common ground or be uncertain about the human's higher level intents in their approach. \new{For example, an object the Commander determines to be of interest from a static photo may be localized on the dynamic 2D map, and used in the future to resolve references and assist when grounding single utterances as annotated with Standard-AMR and Dialogue-AMR.} These potential strategies 
should ultimately be incorporated into the computational approach.

\section{Related Work}
\label{sec:related-work}
This research is at the intersection of NLP--including semantic parsing and dialogue systems--and robotics. 
We limit our direct comparison here to similarly interdisciplinary work; see \cite{tellex2020robots} for a 
full review of research in robotics and language. 
Outside of the work on the grounding approach 
that we directly augment for AMR \citep{howard2021intelligence}, field 
robotics has largely focused on robots that receive an initial, static tasking and then operate autonomously (e.g., \cite{williams2012monitoring,arvidson2010spirit,camilli2010tracking,chiou2022robot}), or on robots that 
are tele-operated (e.g., \cite{kang2003robhaz,ryu2004multi,yamauchi2004packbot}). In contrast, there is relatively little work like ours, seeking to develop robots that are able to be tasked dynamically and interactively via natural language.  

There are, however, a few notable exceptions. \cite{walter2015situationally} describe the development of a voice-controlled fork lift. In contrast to our own research, however, the natural language instructions are more constrained to particular hard-coded commands mentioning a more limited range of objects that are classified in the robot's world model.  Additionally, \cite{heikkila2012affordance} develop a mobile manipulator 
designed for space operations that is capable of accepting spoken commands. This work was similarly focused on a set of domain-specific tasks, but does allow for general spoken commands including following and stopping. It does not, however, allow for bi-directional dialogue with the user.
Unlike both of the previously mentioned voice-controlled robots, it is important to note that our architecture aims to support bi-directional communication between the robot and the Commander, so that ambiguities that might arise in changing environments can be resolved. 

There is also relevant research leveraging LLMs
map or translate between unconstrained natural language and the controlled planning languages of agents more broadly. \cite{song2022one} utilize GPT for deciding upon the appropriate high-level plan given natural language instructions, and then use a more traditional low-level planning component to execute specific motor movements 
to grounded points in the environment. 
The high-level and low-level models are also able to communicate, such that the high-level model can be queried for new and updated plans if conflicts arise in the low-level planning model. The plans are multi-step and involve common kitchen interactions and manipulations, including using a knife to slice an apple and heating a potato in the microwave \citep{shridhar2020alfred}.
\cite{palme2023} develop their own multi-modal ``embodied'' language model, called PaLM-E, which accepts both sensor data, such as image data, and natural language text. The model outputs text data that can be interpreted as robot policies. In general, we see potential for leveraging language models in the future both for providing some \textit{a priori}, zero-shot knowledge of objects that the robot might encounter in its environment, which can be used to inform the interpretation of natural language instructions, as well as for providing a likely mapping between unconstrained natural language and the constrained set of robot behaviors.  

\new{While there has been a veritable explosion of research leveraging LLMs in robotics, we have not seen broad adoption of LLMs in field robotics, where multiple robots engage with humans in physically situated, complex tasks that are dynamic and potentially dangerous. In such high stakes applications, two major weaknesses of LLMs remain problematic: factual inaccuracies and what is often termed ``hallucination'' of irrelevant information, as well as infeasible and sub-optimal planning. \citet{ren2023robots} provide a path to remedy the first hallucination issue by enabling robots to engage in dialogue and ask for help with the use of a framework for measuring and aligning the uncertainty of LLM-based planners. The limitation of this work is that it assumes fully grounded, known environments and objects, and task feasibility of the plans output by the LLM.  Neither assumption is valid in disaster relief domains.} 

\new{\cite{rana2023sayplan} and others have attempted to remedy infeasible and sub-optimal planning by including a traditional planning step. The authors develop SayPlan, which leverages a LLM along with traditional planner in order to enable natural language instruction input to a manipulator arm robot in household and office tasks. The research relies upon a pre-populated 3D scene graph. Innovations include using the LLM to search over and condense the scene graph to only the relevant portion of the graph for the plan (called ``semantic search''), which keeps the input to the in context learning window small. Additionally, the research leverages iterative replanning (called ``causal planning'') to ensure executable plans, leveraging chain of thought prompting and testing each plan against the condensed scene graph. While the authors obtain promising results, the challenges of the system include: negation (e.g., \textit{find the office with no/without cabinets}), counting (e.g., \textit{which office has more than one t-shirt}), and spatial reasoning (e.g., \textit{find the office closest to the entrance}). We note that our grounding approach readily handles spatial reasoning and is equipped to deal with negation, given the consistent and explicit representation of negation in AMR.}

Finally, explainability is critical for adoption of robotic systems in high-stakes tasks such as disaster relief; thus, further research enabling transparency and explainability of systems leveraging LLMs is needed.
Neuro-symbolic approaches (e.g., \cite{RichEmbeddings2022}) are promising for providing greater transparency. 
For example, \cite{zhang2022danli} develop 
DANLI, which symbolically represents subgoals as predicates on objects in the robot’s world model. 

There is a growing body of research leveraging AMR for NLU in human-agent interaction. The present research is part of a broader ongoing research effort leveraging a two-step NLU pipeline that first parses natural language into AMR, which abstracts away from some surface variation, but then in a second step converts the Standard-AMR into Dialogue-AMR. 
While the present research systems leverage Standard-AMR as the input to the grounding component, we will shift to using Dialogue-AMR as the input parse, as we expect that the further normalization will allow us to achieve comparable results with even less training data. Furthermore, Dialogue-AMR leverages spatial rolesets from Spatial-AMR \citep{bonn2020spatial}, which provides detailed relations for spatial relations for expressions such as \textit{in front of}, which currently does not have a detailed representation with a relational concept in Standard-AMR.  

Other research to augment AMR for interaction includes work to further develop multi-modal, gestural AMR \citep{brutti2022abstract} as well as efforts to further develop aspect and modality in AMR to support NLU \citep{donatelli-etal-2020-two}. Finally, there is research in leveraging AMR parses of image captions in order to develop scene graphs, which can help agents to summarize and process visual scenes (e.g., \cite{choi2022scene} and \cite{choi2022sgram}.) 
Together, all of these threads of research demonstrate ways in which AMR can serve as a unified representation for making sense of multiple modalities of information.  

An ongoing research question that we have touched upon is how to incorporate the multi-modal information into the computational pipeline. Before that decision can even be addressed, the overlap and contribution of each modality must be established within the experimental context. Using individual objects within a laboratory setting, e.g., an apple on a plate, \cite{kebe2021spoken} record RGB images and depth point clouds of physical objects, then crowdsource textual and spoken descriptions about the objects. In this way, the physical object apple has four different representations each from a unique modality. This work has shown that when one input stream is degraded or unavailable, algorithms using the other modalities can still identify the target object. Our in-situ problem space may complicate the collection of `clean' data pertaining to each object, although such `clean' representations may be beneficial for approximating a best guess, for example, if the Commander called an object a \textit{shredder}, but the image, even though it might be dark, and LIDAR signified it was a \textit{suitcase}. Another approach called ConceptFusion combines in-situ multi-modal data from RGB images and LIDAR to construct a 3D map, which is then queried at specific locations within the map for textual, audio, and click information, e.g., a chair is described as `A comfy place to sit and watch tv' \citep{jatavallabhula2023conceptfusion}. While the goal of this work is to jointly process all the modalities, there are key differences from our problem space. First, our robot already has all the visual information about the environment. As described in Section~\ref{ssec:visual}, the live video and 3D LIDAR cannot be shared with the human as it would violate the low-bandwidth constraint, so even if the robot had this 3D map, the human couldn't directly interact with it. A second difference is that our Commanders build up an understanding of the space over time as they explore, rather than all at once.

\section{Conclusions \& Future Work}
\label{sec:conclusion}

This research brings together several layers of annotation on a multi-modal corpus of human-robot dialogue in order to support effective task-based dialogue in mixed human-robot teams. Each layer of annotation supports establishing and maintaining common ground between human and robot interlocutors, which is critical for communicating effectively about the physical environment, given that the human has limited understanding of that environment. 

Several broad challenges remain. The first is an architectural challenge---current architectural solutions do not support the level of cross-modal, bi-directional communication that is needed to overcome miscommunications about the environment and reason over natural language instructions before, during, and after grounding, planning, and execution of a task. Related to this architectural challenge is the challenge of how a human can effectively interface with one or more robots in a remote environment. Our research into image request strategies and annotation of LIDAR maps has begun to reveal the potential of this visual information for conveying the robot's model of the environment, which is critical for establishing and maintaining common ground.  Humans and robots experience the world in different ways, so how can we find ways of visualizing and modeling the robot's environment in a way that is understandable to people?  We continue to work towards a system that dynamically and continuously allows for cross-modal, bi-directional communication among the components of that system, as well as an interface that effectively conveys the system status, including disparate sensor types, and environment to the operator.  

\backmatter

\bmhead{Funding} 
This work was supported in part by the U.S. Army Research
Laboratory under the Advanced Research Technology, Inc. contract numbers W911QX-18-F-0096 and W911NF-14-D-0005.

\bmhead{Data availability} The SCOUT dataset is available \new{on github, and contains the transcripts, AMR and Dialogue-AMR annotations, Dialogue Structure annotations, images, maps, and annotations. \url{https://github.com/USArmyResearchLab/ARL-SCOUT}}

\bmhead{Author contribution} All authors contributed to the development and/or annotation of one or more of the schemas in this work.

\section*{Declarations}

\bmhead{Conflict of interest} The authors declare that they have no conflict of interest.

\bibliography{LREVBotLAnnots}

\begin{appendices}

\section{}\label{sec:appendix}

Table~\ref{tab:amr-ds} shows Excerpt \hyperref[excerpt2]{2} as annotated with Dialogue-AMR and Dialogue Structure (TU, antecedent, relations) annotations. Table~\ref{tab:RobotLexicon} shows the robot concepts with associated Dialogue-AMR relations, compatible speech act types, and examples. 

\begin{table*}[ht!]
\caption{\label{tab:amr-ds} \new{Excerpt \hyperref[excerpt2]{2}} shown with Dialogue-AMR and Dialogue Structure Annotations} 
\adjustbox{max width=1\textwidth}{%

\begin{tabular}{lp{1.8in}p{0.2in}p{0.15in}p{0.65in}p{3.0in}}

\toprule                

\# & {\bf Utterance} & {\bf TU} & {\bf Ant}& {\bf Rel} & {\bf Dialogue-AMR} \\ 
\midrule
  
1 & CMD: robot continue down the hallway directly in front of you underneath the overhead light & 1 & & & \begin{verbatim}(c / command-00
 :ARG0 (c3 / commander)
 :ARG1 (g / go-01 :completable -
   :ARG0 r
   :path (h / hallway
     :ARG1-of (f / front-03
       :ARG2 r)
     :location (u / underneath
       :op1 (l / light
         :mod (o / overhead))))
     :time (a / after
       :op1 (n2 / now)))
  :ARG2 (r / robot))
\end{verbatim}
\\ \hline
\rowcolor{gray} 2 & DM$\rightarrow$CMD: I don't see an overhead light in my current position. Would you like me to send a photo? & 1 & 1 & offer & \begin{verbatim}(a / assert-00
 :ARG0 (r / robot-dm)
 :ARG1 (s / see-01 :polarity - :stable -
   :ARG0 r
   :ARG1 (l / light
      :mod (o / overhead))
      :time (n / now))
 :ARG2 (c / commander))
(o / offer-00
 :ARG0 (r / robot-dm)
 :ARG1 (s / send-image-99 :completable +
   :ARG0 r
   :ARG1 (i2 / in-front-of
     :op1 r)
   :ARG2 c
   :time (a / after
     :op1 (n / now)))
:ARG2 (c / commander))
\end{verbatim}
\\ \hline
3 & CMD: robot send a photo & 1 & 2 & offer-accept & \begin{verbatim}(c / command-00
 :ARG0 (c2 / commander-dm)
 :ARG1 (s / send-image-99 :completable +
   :ARG0 r
   :ARG1 (i2 / in-front-of
     :op1 r)
   :ARG2 c2
   :time (a3 / after
     :op1 (n2 / now)))
 :ARG2 (r / robot-rn))
 \end{verbatim}
\\ \hline
\rowcolor{gray} 4 & DM$\rightarrow$RN: photo & 1 & 3 & translation-r-direct & \begin{verbatim}(c / command-00
 :ARG0 (c2 / commander-dm)
 :ARG1 (s / send-image-99 :completable +
   :ARG0 r
   :ARG1 (i2 / in-front-of
     :op1 r)
   :ARG2 c2
   :time (a3 / after
     :op1 (n2 / now)))
 :ARG2 (r / robot-rn))
\end{verbatim}
\\ \hline
\end{tabular}}
\end{table*}

\begin{table*}[h!]
\centering
\caption{Robot concepts with associated Dialogue-AMR relations, compatible speech act types, and examples. Within the annotated corpus release, the speech act labels shown in this table are marked with the ending ``-SA'' (e.g., \amr{Question-SA, Assertion-SA}, etc.). \label{tab:RobotLexicon}}
\begin{adjustbox}{width=\textwidth}
\begin{tabular}{|l|l|l|l|}
\hline
\textbf{Robot Concepts} & \textbf{\begin{tabular}[c]{@{}l@{}}Dialogue-AMR\\ Relations\end{tabular}}  & \textbf{\begin{tabular}[c]{@{}l@{}}Compatible\\ Speech Acts\end{tabular}}                           & \textbf{Examples}                                                                                                                                          \\ \hline
\textsc{Ability}                & Able-01                                                                   & \begin{tabular}[c]{@{}l@{}}Question, \\Assertion\end{tabular}                                      & \textit{\begin{tabular}[c]{@{}l@{}}Are you able to move that orange cone in front of you?; \\ I'm not able to manipulate objects.\end{tabular}}            \\ \hline
\textsc{Scene}                  & See-01                                                                    & \begin{tabular}[c]{@{}l@{}}Question,\\ Assertion\end{tabular}                                       & \textit{\begin{tabular}[c]{@{}l@{}}Do you see foreign writing?; \\ I see two yellow helmets to my left.\end{tabular}}                                     \\ \hline
\textsc{Environment}            & Sense-01                                                                  & \begin{tabular}[c]{@{}l@{}}Question, \\ Assertion\end{tabular}                                      & \textit{\begin{tabular}[c]{@{}l@{}}What is the current temperature?;\\ My LIDAR map is showing no space behind the TV.\end{tabular}}                      \\ \hline
\textsc{Readiness}              & Ready-02                                                                  & \begin{tabular}[c]{@{}l@{}}Question, \\ Assertion\end{tabular}                                      & \textit{\begin{tabular}[c]{@{}l@{}}Are you ready?; \\ I'm ready.\end{tabular}}                                                                            \\ \hline
\textsc{Familiarity}            & Familiarize-01                                                            & \begin{tabular}[c]{@{}l@{}}  Assertion, \\ Open-Option\end{tabular}                                      & \textit{\begin{tabular}[c]{@{}l@{}}I think you are more familiar with shoes than I am; \\ If you describe an object, you can help me learn what it is.\end{tabular}}                                                                                              \\ \hline
\textsc{Equipment}             & Equip-01                                                                  & \begin{tabular}[c]{@{}l@{}}Question, \\ Assertion\end{tabular}                                      & \textit{\begin{tabular}[c]{@{}l@{}}What kind of sensors do you have?; \\ I have no arms, only wheels!\end{tabular}}                                       \\ \hline
\textsc{Memory}                 & Remember-01                                                               & \begin{tabular}[c]{@{}l@{}}Question, \\ Assertion\end{tabular}                                      & \textit{\begin{tabular}[c]{@{}l@{}}How did we get here from last time?;\\ Yes (we've been here before).\end{tabular}}                                         \\ \hline
\textsc{Processing}             & Process-01                                                                & Assertion                                                                                           & \textit{\begin{tabular}[c]{@{}l@{}}Processing...; \\ Hmm...\end{tabular}}                                                                                 \\ \hline
\textsc{Task}                   & Task-01                                                                   & \begin{tabular}[c]{@{}l@{}} Assertion, \\ Command\end{tabular}                           & \textit{\begin{tabular}[c]{@{}l@{}}We're looking for doorways; \\ End task. \end{tabular}}                                                                 \\ \hline
\textsc{Send-Image}             & \begin{tabular}[c]{@{}l@{}}Send-image-XX\\ (domain-specific)\end{tabular} & \begin{tabular}[c]{@{}l@{}}Assertion, \\ Offer, \\ Command, \\ Open-Option, \\ Promise\end{tabular} & \textit{\begin{tabular}[c]{@{}l@{}}Image sent; \\ Would you like me to take a picture? \\ Take a picture; \\ I can send a picture; \\ I will send a picture.\end{tabular}}                                                                           \\ \hline
\textsc{Movement}               & Go-02                                                                     & \begin{tabular}[c]{@{}l@{}}Assertion, \\ Offer, \\ Command, \\ Open-Option, \\ Promise\end{tabular} & \textit{\begin{tabular}[c]{@{}l@{}}I moved forward one foot\\ I will move forward one foot, ok? \\ Back up three feet; \\ You can tell me to move a certain distance or to move to an object; \\I will move forward one foot.\end{tabular}}                                      \\ \hline
\textsc{Rotation}               & Turn-01                                                                   & \begin{tabular}[c]{@{}l@{}}Assertion, \\ Command, \\ Open-Option, \\ Promise\end{tabular} & \textit{\begin{tabular}[c]{@{}l@{}}Turning...\\ Turn to face West; \\ You can tell me to turn a number of degrees or to face something; \\ I will turn 90 degrees.\end{tabular}}                                               \\ \hline
\textsc{Repeat}                 & Repeat-01                                                                 & \begin{tabular}[c]{@{}l@{}}Offer, \\ Command, \\ Request\end{tabular} & \textit{\begin{tabular}[c]{@{}l@{}}Would you like me to repeat the last action?; \\ Do the following four times... \\ Can you repeat that?\end{tabular}} \\ \hline
\textsc{Cancel}                 & Cancel-01                                                                 & \begin{tabular}[c]{@{}l@{}}Command\end{tabular}                                       & \textit{\begin{tabular}[c]{@{}l@{}}Cancel command; Stop; Nevermind\end{tabular}}                                                                    \\ \hline
\textsc{Do}                     & Do-02                                                                     & \begin{tabular}[c]{@{}l@{}}Question, \\ Assertion\end{tabular}                                      & \textit{\begin{tabular}[c]{@{}l@{}}Did I successfully do what you asked? \\ Executing; Done\end{tabular}}                                                                                       \\ \hline
\textsc{Clarify}                & Clarify-10                                                                & \begin{tabular}[c]{@{}l@{}}Assertion, \\ Request\end{tabular}                                       & \textit{\begin{tabular}[c]{@{}l@{}}Brown, not round; \\ How much is a little bit?\end{tabular}}                      \\ \hline
\textsc{Stop} (motion)          & Stop-01                                                                   & \begin{tabular}[c]{@{}l@{}}Command\end{tabular}                                       & \textit{\begin{tabular}[c]{@{}l@{}}Stop there; Stop!\end{tabular}}                                                                                     \\ \hline
\textsc{Help}                   & Help-01                                                                   & \begin{tabular}[c]{@{}l@{}}Command, \\ Request, \\ Open-Option\end{tabular}               & \textit{\begin{tabular}[c]{@{}l@{}}Help!\\ I need your help to find shoes;\\ You can ask for help at any time.\end{tabular}}                                                                 \\ \hline
\textsc{Locate}                 & Locate-02                                                                 & \begin{tabular}[c]{@{}l@{}}Assertion, \\ Command\end{tabular}                                       & \textit{\begin{tabular}[c]{@{}l@{}}(I've located) 3; \\ Find doorways; ...and locate shoes\end{tabular}}                                                                    \\ \hline
\textsc{Calibrate}              & Calibrate-01                                                              & \begin{tabular}[c]{@{}l@{}}Assertion, \\ Command\end{tabular}                                       & \textit{\begin{tabular}[c]{@{}l@{}} Calibrating...; Calibration complete\\ Calibrate\end{tabular}}                                                         \\ \hline
\textsc{Instruct}               & Instruct-01                                                               & Request                                                                                             & \textit{\begin{tabular}[c]{@{}l@{}}What should we do next?; Then what?\end{tabular}}                                                                     \\ \hline
\textsc{Wait}                  & Wait-01                                                                   & \begin{tabular}[c]{@{}l@{}}Command, \\ Request\end{tabular}                           & \textit{\begin{tabular}[c]{@{}l@{}}Wait!\\ Please wait.\end{tabular}}                                                      \\ \hline
\textsc{Permission}             & Permit-01                                                                 & \begin{tabular}[c]{@{}l@{}} Request\end{tabular}                                       & \textit{Robot, can I call you Fido?}                                                                                                                      \\ \hline
\textsc{Understanding}             & Understand-01                                                                  & \begin{tabular}[c]{@{}l@{}}Question, \\ Assertion

\end{tabular}                                      & \textit{\begin{tabular}[c]{@{}l@{}}Did I misunderstand?; \\ Ok, I think I got it.\end{tabular}}                                       \\ \hline
\end{tabular}
\end{adjustbox}
\end{table*}




\end{appendices}



\end{document}